# The Gravity of the Classical Klein-Gordon field

Piero Chiarelli

*National Council of Research of Italy, Moruzzi 1, 56124 Pisa, Italy*

*Interdepartmental Center "E. Piaggio" University of Pisa, Diotisalvi, 2, 56122 Pisa, Italy*

Phone: +39-050-315-2359
Fax: +39-050-315-2166

Email: pchiare@ifc.cnr.it.

**Abstract:** The work shows that the evolution of the field of the free Klein-Gordon equation (KGE), in the hydrodynamic representation, can be represented by the motion of a mass density $\propto|\cdot|^2$ subject to the Bohm-type quantum potential, whose equation can be derived by a minimum action principle. Once the quantum hydrodynamic motion equations have been covariantly extended to the curved space-time, the gravity equation (GE), determining the geometry of the space-time, is obtained by minimizing the overall action comprehending the gravitational field. The derived Einstein-like gravity for the KGE field shows an energy-impulse tensor density (EITD) that is a function of the field with the spontaneous emergence of the "cosmological" pressure tensor density (CPTD) that in the classical limit leads to the cosmological constant (CC). The energy-impulse tensor of the theory shows analogies with the modified Brans-Dick gravity with an effective gravity constant G divided by the field squared. Even if the classical cosmological constant is set to zero, the model shows the emergence of a theory-derived quantum CPTD that, in principle, allows to have a stable quantum vacuum (out of the *collapsed branched polymer phase*) without postulating a non-zero classical CC. In the classical macroscopic limit, the gravity equation of the KGE field leads to the Einstein equation. Moreover, if the boson field of the photon is considered, the EITD correctly leads to its electromagnetic energy-impulse tensor density.

The work shows that the cosmological constant can be considered as a second order correction to the Newtonian gravity. The outputs of the theory show that the expectation value of the CPTD is independent by the zero-point vacuum energy density and that it takes contribution only from the space where the mass is localized (and the space-time is curvilinear) while tends to zero as the space-time approaches to the flat vacuum, leading to an overall cosmological effect on the motion of the galaxies that may possibly be compatible with the astronomical observations.

## 1. Introduction

One of the serious problems the nowadays gravity physics [1] refers to the connection between the quantum fields theory (QFT) and the GE. The problem has come to a partial solution in the semi-classical approximation where the energy-impulse tensor density is substituted by its expectation value [2-6].

Even if unable to give answers in a fully quantum regime, the semiclassical approximation has brought to successful results such as the explanation of the Hawking radiation and BH evaporation [7].

The difficulties about the integration of QFT and the GE become really evident in the so called cosmological constant problem, a term that Einstein added to its equation to give stability to the solution of universe evolution that in the general relativity would lead to its final collapse. The introduction, by hand, of the cosmological constant was then refused by Einstein himself that defined it as *the biggest mistake of my life* [8]. Actually, the CC has been introduced [1] to explain the astronomical observations about the motion of galaxies [9] and to give stability to the quantum vacuum bringing it out from the unphysical *collapsed branched polymer phase* [10], being the physical vacuum of the *strong gravity phase* related to a positive, not null, CC [11]. Moreover, the EITD for classical bodies in the GE owns a point-dependence by the mass density (i.e., $\propto|\psi|^2$), without any analytical complete connection with the field of matter $\psi=|\psi|exp\frac{i}{\hbar}S$. As discussed by Thiemann [12], this connection cannot be build up by simply replacing the EITD by its Minkowskian vacuum expectation value. If we do so, we end with a non-Minkowskian metric tensor solution that has to feed back into the vacuum expectation value and so on with the iteration that does not converge in general.

As a consequence of this fault, modifications to the Einstein equation have been proposed both from theoretical point of view, such as the Brans-Dicke modified gravity [13], and by using a semi-empirical approach such as the *covariant running G* or the *slip function* QFT [14]; However, the cosmological constant itself in the quantum pure gravity can be considered a modification of the general relativity for defining a GE compatible with the needs of a quantum theory.

Due to the undefined connection between the GE and the particle fields, the integration between the QFT and the GE is still an open question that is object of intense theoretical investigation. Generally speaking, the link between the matter fields and the GE can be obtained by:

1. Defining an adequate GE for matter fields (as, for instance, happens for the photon field);
2. Defining the link between the GE and the QFT by quantizing the action of the new GE.

At glance with the first point, the paper shows that it is possible to obtain the GE with analytical connection with the KGE field

$$\psi=|\psi|exp\frac{i}{\hbar}S. \qquad (1.0.1)$$

To this end, the hydrodynamic representation of the field equation as a function of the variables

$$|\psi| \text{ and } \partial_\mu S=-p_\mu \qquad (1.0.2)$$

(that leads to the classical-like description of a mass density $|\psi|^2$ owing the hydrodynamic impulse $p_\mu$. subject to the non-local quantum potential interaction) is utilized. Then, by using the minimum action principle applied to the hydrodynamic model, the gravity generated by the KGE field $\psi$ is derived.

The paper is organized as follows: In section 2 the Lagrangean version of the hydrodynamic KGE is developed; In section 3 the gravity equation is derived by the minimum action principle; In section 4, the perturbative approach to the GE-KGE system of evolutionary equations is derived; In section 5, the expectation value of the cosmological constant of the quantum KGE massive field is calculated; In section 6, some features of the GE as well as the check of the theory are discussed.

## 2. The hydrodynamic representation of the Klein Gordon equation

In this section we derive the hydrodynamic representation of the Klein-Gordon equation (KGE) in the form of Lagrangean equations that allow to define the minimum action principle for the hydrodynamic formalism.

Following the method firstly proposed by Madelung [15] and then generalized by other authors [16-18], the hydrodynamic form of the KGE

$$\partial_\mu\partial^\mu\psi=-\frac{m^2c^2}{\hbar^2}\psi, \qquad (2.0.1)$$

in the Minkowskian space, is given by the system of two differential equations [19: The Hamilton-Jacobi type one

$$\partial_\mu S\ \partial^\mu S-\hbar^2\frac{\partial_\mu\partial^\mu|\psi|}{|\psi|}-m^2c^2=0 \qquad (2.0.2)$$

coupled to the current conservation equation

$$\frac{\partial}{\partial q_«}\left( | \cdot |^2 \frac{\partial S}{\partial q^«} \right) = m \frac{\partial J_«}{\partial q_«} = 0 \qquad (2.0.3)$$

where

$$S = \frac{\hbar}{2i} ln[\frac{\cdot}{\cdot\ *}] \qquad (2.0.4)$$

and where the 4-current reads

$$J_« = (c° , -J_i) = \frac{i\hbar}{2m}( \cdot\ * \frac{\partial \cdot}{\partial q^«} - \cdot\ \frac{\partial \cdot\ *}{\partial q^«} ). \qquad (2.0.5)$$

Moreover, being the 4-impulse in the hydrodynamic analogy

$$p_« = (\frac{E}{c}, -p_i) = -\frac{\partial S}{\partial q^«} \qquad (2.0.6)$$

it follows that

$$J_« = (c° , -J_i) = -| \cdot |^2 \frac{p_«}{m} \qquad (2.0.7)$$

where

$$° = \frac{J_0}{c} = \frac{| \cdot |^2}{mc^2} \frac{\partial S}{\partial t} \qquad (2.0.8)$$

Moreover, by using (2.0.6), equation (2.0.2) reads

$$\frac{\partial S}{\partial q^«} \frac{\partial S}{\partial q_«} = p_« p^« = \left( \frac{E^2}{c^2} - p^2 \right) = m^2 c^2 \left( 1 - \frac{V_{qu}}{mc^2} \right) \qquad (2.0.9)$$

(where $p^2 = p_i p_i$ is the modulus squared of the hydrodynamic spatial momentum (Italic indexes run from 1 to 3)) that, for k-th eigenstate $\cdot_k = |\cdot_k| exp \frac{i}{\hbar} S_{(k)}$, leads to

$$\begin{aligned} \frac{E_{(k)}{}^2}{c^2} - p_{(k)}{}^2 &= m^2 ¥^2 \dot{q}_« \dot{q}^« \left( 1 - \frac{V_{qu(k)}}{mc^2} \right) \\ &= m^2 ¥^2\ c^2 \left( 1 - \frac{V_{qu(k)}}{mc^2} \right) - m^2 ¥^2\ \dot{q}^2 \left( 1 - \frac{V_{qu(k)}}{mc^2} \right) \end{aligned} \qquad (2.0.10)$$

where

$$¥ = \frac{1}{\sqrt{1 - \frac{\dot{q}^2}{c^2}}} = \frac{1}{\sqrt{\frac{g_{«¬} \dot{q}^¬ \dot{q}^«}{c^2}}}. \qquad (2.0.11)$$

Moreover, by denoting the negative-energy state by the minus subscript, so that $E_{(k)\pm} = \pm E_{(k)}$, from (2.0.10) it follows that

$$E_{(k)} = m\gamma c^2 \sqrt{1 - \frac{V_{qu(k)}}{mc^2}} \qquad (2.0.12)$$

and more generally, by defining $p_{(k)\mu\pm} = \pm p_{(k)\mu}$, the hydrodynamic impulse [19] reads

$$p_{(k)\mu} = m\gamma_{(k)}\ \dot{q}_{(k)\mu} \sqrt{1 - \frac{V_{qu(k)}}{mc^2}} = \frac{E_{(k)}}{c^2} \dot{q}_{(k)\mu} \qquad (2.0.13)$$

where $V_{qu(k)} = V_{qu(\psi_k)}$ is the quantum potential that reads

$$V_{qu(\psi)} = -\frac{\hbar^2}{m} \frac{\partial_\mu \partial^\mu |\psi|}{|\psi|}. \qquad (2.0.14)$$

**2.1 The Lagrangean form of the KGE**

Equation (2.02) in the low velocity limit leads to the Madelung quantum hydrodynamic analogy of Schrodinger equation [19] that in the classical limit (i.e., $\hbar = 0$ ) leads to the classical Lagrangean equation of motion [20]. For the purpose of this work, we generalize the Lagrangean formulation to the hydrodynamic KGEs (2.02-3).

Since in curved space-time under the gravitational field we may have discrete energy values, in the following we use the discrete formalism (i.e., $\iint \frac{d^3k}{(2\pi)^3} \to \frac{1}{\sqrt{V}} \sum_{k=-\infty}^{\infty}$ ) that is also useful in the numerical approach.

For the generic superposition of eigenstates

$$\psi = \sum_{k=k_{min}}^{k=k_{max}} b_k |\psi_k| exp[\frac{iS_{(k)}}{\hbar}] = |\psi| exp[\frac{iS}{\hbar}], \qquad (2.1.1)$$

where

$$S = \frac{\hbar}{2i} ln[\frac{\psi}{\psi^*}] = \frac{\hbar}{2i} \begin{pmatrix} ln[\sum_{k=k_{min}}^{k=k_{max}} b_k |\psi_k| exp(\frac{iS_{(k)}}{\hbar})] \\ -ln[\sum_{k=k_{min}}^{k=k_{max}} b^*_k |\psi_k| exp(-\frac{iS_{(k)}}{\hbar})] \end{pmatrix}, \qquad (2.1.2)$$

by using (2.0.4, 2.0.6), it follows that the hydrodynamic momentum $p_\mu$, and the Lagrangean function, respectively, read

$$p_{«} = -\partial_{«} S = -\frac{1}{2}\frac{\sum_{k=k_{min}}^{k=k_{max}} b_k \,|\cdot_k| exp[\frac{iS_{(k)}}{\hbar}]\left(\frac{\hbar}{i}\partial_{«}\, ln|\cdot_k| - p_{(k)«}\right)}{\sum_k b_k \,|\cdot_k| exp[\frac{iS_{(k)}}{\hbar}]} + \frac{1}{2}\frac{\sum_{k=k_{min}}^{k=k_{max}} b*_k \,|\cdot_k| exp[\frac{-iS_{(k)}}{\hbar}]\left(\frac{\hbar}{i}\partial_{«}\, ln|\cdot_k| + p_{(k)«}\right)}{\sum_k b*_k \,|\cdot_k| exp[\frac{-iS_{(k)}}{\hbar}]} = Tr\left(\boldsymbol{p}_{«}\right), \quad (2.1.3)$$

where the matrix $\boldsymbol{p}_{«}$ reads

$$\boldsymbol{p}_{«} \equiv \left(\boldsymbol{p}_{«}\right)_{jk} = \left( -\frac{1}{2}\frac{b_k \,|\cdot_k| exp[\frac{iS_{(k)}}{\hbar}]\left(\frac{\hbar}{i}\partial_{«}\, ln|\cdot_k| - p_{(k)«}\right)}{\sum_j b_j \,|\cdot_j| exp[\frac{iS_{(j)}}{\hbar}]} + \frac{1}{2}\frac{b*_k \,|\cdot_k| exp[\frac{-iS_{(k)}}{\hbar}]\left(\frac{\hbar}{i}\partial_{«}\, ln|\cdot_k| + p_{(k)«}\right)}{\sum_j b*_j \,|\cdot_j| exp[\frac{-iS_{(j)}}{\hbar}]} \right) ¢_{jk} = \tilde{p}_{(k)«} ¢_{jk} \quad (2.1.4)$$

and where *j, k* run from $k_{min}$ and $k_{max}$;

$$L = \frac{dS}{dt} = \frac{\partial S}{\partial t} + \frac{\partial S}{\partial q_i}\dot{q}_i = -Tr\left(\boldsymbol{p}_{«}\dot{\boldsymbol{q}}^{«}\right) = Tr\left(\boldsymbol{L}\right)$$

$$= \frac{1}{2}\frac{\sum_{k=-\infty} b_k \,|\cdot_k| exp[\frac{iS_{(k)}}{\hbar}]\left(\frac{\hbar}{i}\dot{q}^{«}_{(k)}\,\partial_{«}\, ln|\cdot_k| + L_{(k)}\right)}{\sum_k b_k \,|\cdot_k| exp[\frac{iS_{(k)}}{\hbar}]} - \frac{1}{2}\frac{\sum_{k=-\infty} b*_k \,|\cdot_k| exp[\frac{-iS_{(k)}}{\hbar}]\left(\frac{\hbar}{i}\dot{q}^{«}_{(k)}\partial_{«}\, ln|\cdot_k| - L_{(k)}\right)}{\sum_k b*_k \,|\cdot_k| exp[\frac{-iS_{(k)}}{\hbar}]}. \quad (2.1.5)$$

where

$$\boldsymbol{L} = -\boldsymbol{p}_{«}\dot{\boldsymbol{q}}^{«} = -\tilde{p}_{(h)«}\dot{q}^{«}_{(k)}¢_{hk} = \tilde{L}_{(k)}¢_{hk} \quad (2.1.6)$$

where

$$\tilde{L}_{(k)} = \frac{1}{2}\left( \begin{array}{l} \dfrac{b_k \,|\cdot_k|\, exp[\frac{iS_{(k)}}{\hbar}] \left( \frac{\hbar}{i} \dot{q}^{«}_{(k)} \,\partial_{«}\, ln|\cdot_k| + L_{(k)} \right)}{\sum_k b_k \,|\cdot_k|\, exp[\frac{iS_{(k)}}{\hbar}]} \\ - \dfrac{b*_k \,|\cdot_k|\, exp[\frac{-iS_{(k)}}{\hbar}] \left( \frac{\hbar}{i} \dot{q}^{«}_{(k)} \partial_{«}\, ln|\cdot_k| - L_{(k)} \right)}{\sum_k b*_k \,|\cdot_k|\, exp[\frac{-iS_{(k)}}{\hbar}]} \end{array} \right) \quad (2.1.7)$$

$$\dot{\boldsymbol{q}}^{«} = \dot{q}^{«}_{(k)} ¢_{jk}\,; \quad (2.1.8)$$

Thence, by using the identities $L_{(k)} = -p_{(k)«}\, \dot{q}^{«}_{(k)}$ and $\left( \frac{\partial}{\partial \dot{\boldsymbol{q}}^{«}} \right)_{jh} = \left( \frac{\partial}{\partial \dot{q}^{«}_{(j)}} \right) ¢_{jh}$ it follows that

$$\boldsymbol{p}_{«} = \left( \boldsymbol{p}_{«} \right)_{jk} = -\left( \frac{\partial}{\partial \dot{\boldsymbol{q}}^{«}} \right)_{jh} \left( \boldsymbol{L} \right)_{hk} = -\left( \frac{\partial}{\partial \dot{q}^{«}_{(j)}} \right) \tilde{L}_{(k)} ¢_{jk} = -\frac{\partial \boldsymbol{L}}{\partial \dot{\boldsymbol{q}}^{«}} \quad (2.1.9)$$

$$\begin{aligned} \tilde{p}_{(k)«} = & -\frac{1}{2} \frac{b_k \,|\cdot_k|\, exp[\frac{iS_{(k)}}{\hbar}] \left( \frac{\hbar}{i} \partial_{«}\, ln|\cdot_k| + \frac{\partial L_{(k)}}{\partial \dot{q}^{«}} \right)}{\sum_k b_k \,|\cdot_k|\, exp[\frac{iS_{(k)}}{\hbar}]} \\ & + \frac{1}{2} \frac{b*_k \,|\cdot_k|\, exp[\frac{-iS_{(k)}}{\hbar}] \left( \frac{\hbar}{i} \partial_{«}\, ln|\cdot_k| - \frac{\partial L_{(k)}}{\partial \dot{q}^{«}} \right)}{\sum_k b*_k \,|\cdot_k|\, exp[\frac{-iS_{(k)}}{\hbar}]} \\ & = -\frac{\partial \tilde{L}_{(k)}}{\partial \dot{q}^{«}_{(k)}} \end{aligned} \quad (2.1.10)$$

Moreover, given that, for stationary states of time-independent systems (i.e., eigenstates) [19] the hydrodynamic Lagrangean function (2.1.5) does not explicitly depend by time, it holds

$$\tilde{L}_{(k)} = L_{(k)} \quad (2.1.11)$$

$$\tilde{p}_{(k)«} = p_{(k)«} \quad (2.1.12)$$

$$-\frac{\partial L_{(k)}}{\partial q^{«}} = -\frac{\partial}{\partial q^{«}} \frac{dS_{(k)}}{dt} = -\left( 0, -\frac{\partial}{\partial q_i} \frac{dS_{(k)}}{dt} \right) = -\left( 0, -\frac{d}{dt} \frac{\partial S_{(k)}}{\partial q_i} \right) = \frac{d}{dt}\left( -\frac{\partial S_{(k)}}{\partial q^{«}} \right) = \dot{p}_{(k)«}\,, \quad (2.1.13)$$

so that, for the eigenstates, the system of Lagrangean hydrodynamic motion equations read

$$p_{(k)«} = -\frac{\partial L_{(k)}}{\partial \dot{q}^{«}_{(k)}}, \quad (2.1.14)$$

$$\dot{p}_{(k)«} = -\frac{\partial L_{(k)}}{\partial q^{«}} \qquad (2.1.15)$$

where, by using (2.0.13), it follows that

$$L_{(k)} = -\frac{mc^2}{¥_{(k)}}\sqrt{1-\frac{V_{qu(k)}}{mc^2}} = L_{(k)Class}\sqrt{1-\frac{V_{qu(k)}}{mc^2}} = L_{(k)Class} + L_{(k)Q}, \ ( \qquad (2.1.16)$$

where

$$L_{(k)Class} = -\frac{mc^2}{¥_{(k)}}, \qquad (2.1.17)$$

where

$$L_{(k)Q} = -\sim\!\left(V_{qu(k)}\right) L_{(k)Class} = -\left(1-\sqrt{1-\frac{V_{qu(k)}}{mc^2}}\right) L_{(k)Class}, \qquad (2.1.18)$$

where

$$\sim\!\left(V_{qu(k)}\right) = \left(1-\sqrt{1-\frac{V_{qu(k)}}{mc^2}}\right), \qquad (2.1.19)$$

and that

$$\frac{d}{dt}\left(-\frac{\partial L_{(k)}}{\partial \dot{q}^{«}_{(k)}}\right) = -\frac{\partial L_{(k)}}{\partial q^{«}}. \qquad (2.1.20)$$

For sake of accuracy, it must be observed that the solutions of (2.1.20) have to be submitted to quantization (given by the irrotational property [17,19]) and to the current conservation condition (2.0.3) As shown in [19] the stationary states of (2.1.20) obey to the current conservation (2.0.3) and are irrotational solutions (i.e., the eigenstates) for the field of the KGE.

Equation (2.1.11) is the same both for positive and negative energy states since the Lagrangean function $L_{(k)_}$ for negative energy states reads $L_{(k)_} = -p_{(k)«_}\dot{q}^{«}_{(k)} = p_{(k)«}\dot{q}^{«}_{(k)} = -L_{(k)}$

It is useful to observe that the hydrodynamic equation of motion (2.1.20) depends both by $q$, $\dot{q}$ and by the mass distribution $|\cdot|^2$ (and its derivatives) contained in $V_{qu}$.

For $\hbar \to 0$ (i.e.,$V_{qu} \to 0$) the classical motion of the mass distribution for the so-called dust matter [17] are obtained just as a function of $q$, $\dot{q}$.

Moreover, generally speaking, given the integrability of $\tilde{L}_{(k)}$, the motion equation of the generic superposition of state (2.1.1) reads

$$-\frac{\partial \tilde{L}_{(k)}}{\partial q^{«}} = \frac{d}{dt}\left(-\frac{\partial \tilde{L}_{(k)}}{\partial \dot{q}^{«}_{(k)}}\right) + \tilde{p}_{(k)¬}\frac{\partial \dot{q}_{(k)¬}}{\partial q^{«}} = \dot{\tilde{p}}_{(k)«} + \tilde{p}_{(k)¬}\frac{\partial \dot{q}_{(k)¬}}{\partial q^{«}} \qquad (2.1.21)$$

that making the summation over $k$ leads to

$$Tr(\dot{\boldsymbol{p}}_{«}) = -\frac{\partial Tr(\boldsymbol{L})}{\partial q^{«}} - Tr\left(\boldsymbol{p}_{¬}\partial_{«}\dot{\boldsymbol{q}}_{¬}\right) \qquad (2.1.22)$$

and to

$$\dot{p}_{«} = -\frac{\partial L}{\partial q^{«}} - \sum_{k} \tilde{p}_{(k)¬} \frac{\partial \dot{q}_{(k)¬}}{\partial q^{«}} \tag{2.1.23}$$

where

$$Tr\left(\boldsymbol{L}\right) = L = L_{Class} + L_{Q} + L_{mix}, \tag{2.1.24}$$

where

$$L_{Class} = \frac{1}{2} \frac{\sum\limits_{k=-\infty} b_k \,|\cdot_k|\, exp[\frac{iS_{(k)}}{\hbar}]\, L_{(k)Class}}{\sum\limits_{k} b_k \,|\cdot_k|\, exp[\frac{iS_{(k)}}{\hbar}]} + \frac{1}{2} \frac{\sum\limits_{k=-\infty} b^*_k \,|\cdot_k|\, exp[\frac{-iS_{(k)}}{\hbar}]\, L_{(k)Class}}{\sum\limits_{k} b^*_k \,|\cdot_k|\, exp[\frac{-iS_{(k)}}{\hbar}]}, \tag{2.1.25}$$

where

$$L_{Q} = \frac{1}{2} \frac{\sum\limits_{k=-\infty} b_k \,|\cdot_k|\, exp[\frac{iS_{(k)}}{\hbar}]\sim\left(V_{qu(k)}\right) L_{(k)Class}}{\sum\limits_{k} b_k \,|\cdot_k|\, exp[\frac{iS_{(k)}}{\hbar}]} + \frac{1}{2} \frac{\sum\limits_{k=-\infty} b^*_k \,|\cdot_k|\, exp[\frac{-iS_{(k)}}{\hbar}]\sim\left(V_{qu(k)}\right) L_{(k)Class}}{\sum\limits_{k} b^*_k \,|\cdot_k|\, exp[\frac{-iS_{(k)}}{\hbar}]}, \tag{2.1.26}$$

and where

$$L_{mix} = \frac{\hbar}{2i} \frac{\sum\limits_{k=-\infty} b_k \,|\cdot_k|\, exp[\frac{iS_{(k)}}{\hbar}]\, L_{(k)mix}}{\sum\limits_{k} b_k \,|\cdot_k|\, exp[\frac{iS_{(k)}}{\hbar}]} - \frac{\hbar}{2i} \frac{\sum\limits_{k=-\infty} b^*_k \,|\cdot_k|\, exp[\frac{-iS_{(k)}}{\hbar}]\, L_{(k)mix}}{\sum\limits_{k} b^*_k \,|\cdot_k|\, exp[\frac{-iS_{(k)}}{\hbar}]} \tag{2.1.27}$$

where

$$L_{(k)mix} = \dot{q}^{«}_{(k)} \partial_{«} \ln|\cdot_k| \tag{2.1.28}$$

## 2.2 The hydrodynamic energy-impulse tensor

By using the hydrodynamic energy-impulse tensor (EIT) $T_{«}^{¬}$

$$T_{(k)«}^{\;¬} = -\left( \dot{q}^{¬} \frac{\partial L_{(k)}}{\partial \dot{q}^{«}} - L_{(k)} ¢_{«}^{\;¬} \right) = \left( \dot{q}_{(k)}^{¬} p_{(k)«} - \dot{q}_{(k)}^{˜} p_{(k)˜} ¢_{«}^{\;¬} \right)$$

$$= \frac{mc^2}{¥_{(k)}} \sqrt{1 - \frac{V_{qu}}{mc^2}} \left( u_{(k)«} u_{(k)}^{\;¬} - u_{(k)}^{\;˜} u_{(k)˜} ¢_{«}^{\;¬} \right) , \qquad (2.2.1)$$

where $u_{«} = \frac{¥}{c} \dot{q}_{«}$ , with the help of (2.1.10), the quantum hydrodynamic motion equation (2.1.11) reads

$$mc \sqrt{1 - \frac{V_{qu(k)}}{mc^2}} \frac{du_{«}}{dt} = -mcu_{(k)«} \frac{d}{dt} \left( \sqrt{1 - \frac{V_{qu(k)}}{mc^2}} \right) + \frac{mc^2}{¥_{(k)}} \frac{\partial}{\partial q^{«}} \left( \sqrt{1 - \frac{V_{qu(k)}}{mc^2}} \right), \qquad (2.2.2)$$

leading to the equation of motion

$$mc \sqrt{1 - \frac{V_{qu(k)}}{mc^2}} \frac{du_{(k)«}}{dt} = \frac{\partial T_{(k)«}^{\;¬}}{\partial q^{¬}} = T_{(k)«}^{\;¬}{}_{;¬} \, . \qquad (2.2.3)$$

where we pose

$$T_{(k)«}^{\;¬} = T_{(k)Class«}^{\;¬} + L_{(k)Class} ¢_{«}^{\;¬} + T_{(k)Q«}^{\;¬} \qquad (2.2.4)$$

where

$$T_{(k)Class«}^{\;¬} = \frac{mc^2}{¥_{(k)}} u_{«} u^{¬} \qquad (2.2.5)$$

and where

$$T_{(k)Q«}^{\;¬} = -\sim_{\left(V_{qu(k)}\right)} \frac{mc^2}{¥_{(k)}} \left( u_{(k)«} u_{(k)}^{\;¬} - u_{(k)}^{\;˜} u_{(k)˜} ¢_{«}^{\;¬} \right) . \qquad (2.2.6)$$

For the generic quantum state (2.1.1), the energy-impulse tensor reads

$$T_{«}^{\;¬} = -\left( \dot{q}^{¬} \frac{\partial L}{\partial \dot{q}^{«}} - L ¢_{«}^{\;¬} \right) = -\dot{q}^{¬} p_{«} + L ¢_{«}^{\;¬}$$

$$= -\frac{1}{2} \frac{\sum_{k=-\infty} b_k |\cdot_k| exp[\frac{iS_{(k)}}{\hbar}] \left( \frac{\hbar}{i} \left( \dot{q}_{(k)}^{¬} \partial_{«} |\cdot_k| + \dot{q}_{(k)}^{˜} \partial_{˜} |\cdot_k| ¢_{«}^{\;¬} \right) - T_{(k)«}^{\;¬} \right)}{\sum_k b_k |\cdot_k| exp[\frac{iS_{(k)}}{\hbar}]} , \qquad (2.2.7)$$

$$+ \frac{1}{2} \frac{\sum_{k=-\infty} b*_k |\cdot_k| exp[\frac{-iS_{(k)}}{\hbar}] \left( \frac{\hbar}{i} \left( \dot{q}_{(k)}^{¬} \partial_{«} |\cdot_k| + \dot{q}_{(k)}^{˜} \partial_{˜} |\cdot_k| ¢_{«}^{\;¬} \right) + T_{(k)«}^{\;¬} \right)}{\sum_k b*_k |\cdot_k| exp[\frac{-iS_{(k)}}{\hbar}]}$$

it can be recast as

$$T_{«}^{\;¬} = T_{Class«}^{\;¬} + L_{Class} ¢_{«}^{\;¬} + T_{Q«}^{\;¬} + T_{mix«}^{\;¬} \qquad (2.2.8)$$

where

$$
T_{Class\,\mu}{}^{\nu}=\frac{1}{2}\frac{\sum_{k=-\infty} b_k|\psi_k|exp[\frac{iS_{(k)}}{\hbar}]T_{(k)Class\,\mu}{}^{\nu}}{\sum_k b_k|\psi_k|exp[\frac{iS_{(k)}}{\hbar}]}+\frac{1}{2}\frac{\sum_{k=-\infty} b^*_k|\psi_k|exp[\frac{-iS_{(k)}}{\hbar}]T_{(k)Class\,\mu}{}^{\nu}}{\sum_k b^*_k|\psi_k|exp[\frac{-iS_{(k)}}{\hbar}]}, \qquad (2.2.9)
$$

where

$$
T_{Q\mu}{}^{\nu}=\frac{1}{2}\frac{\sum_{k=-\infty} b_k|\psi_k|exp[\frac{iS_{(k)}}{\hbar}]T_{Q(k)\mu}{}^{\nu}}{\sum_k b_k|\psi_k|exp[\frac{iS_{(k)}}{\hbar}]}+\frac{1}{2}\frac{\sum_{k=-\infty} b^*_k|\psi_k|exp[\frac{-iS_{(k)}}{\hbar}]T_{Q(k)\mu}{}^{\nu}}{\sum_k b^*_k|\psi_k|exp[\frac{-iS_{(k)}}{\hbar}]}, \qquad (2.2.10)
$$

and

$$
T_{mix\,\mu}{}^{\nu}=-\frac{\hbar}{2i}\frac{\sum_{k=-\infty} b_k|\psi_k|exp[\frac{iS_{(k)}}{\hbar}]\left(\dot{q}^{\nu}_{(k)}\partial_{\mu}|\psi_k|+\dot{q}^{\lambda}_{(k)}\partial_{\lambda}|\psi_k|\delta_{\mu}{}^{\nu}\right)}{\sum_k b_k|\psi_k|exp[\frac{iS_{(k)}}{\hbar}]}+\frac{\hbar}{2i}\frac{\sum_{k=-\infty} b^*_k|\psi_k|exp[\frac{-iS_{(k)}}{\hbar}]\left(\dot{q}^{\nu}_{(k)}\partial_{\mu}|\psi_k|+\dot{q}^{\lambda}_{(k)}\partial_{\lambda}|\psi_k|\delta_{\mu}{}^{\nu}\right)}{\sum_k b^*_k|\psi_k|exp[\frac{-iS_{(k)}}{\hbar}]}, \qquad (2.2.11)
$$

So far, since we want to covariantly generalize the theory in curvilinear space-time, we have not introduced in the formulas above the explicit form of the Minkowskian KGE field (i.e., $\psi_{k(q_i,t)}=Aexp\left(ik^{\mu}q_{\mu}\right)=Aexp[-i\left(k_iq_i-\omega t\right)]$) with the consequential conditions $\partial^{\nu}|\psi_k|=0$, $k_{min}=-\infty$ and $k_{max}=\infty$. Nevertheless, it is useful to derive such Minkowskian expressions that, by posing

$$
g_{\nu\mu}=\begin{bmatrix}1&0&0&0\\0&-1&0&0\\0&0&-1&0\\0&0&0&-1\end{bmatrix}, \qquad (2.2.12)
$$

read

$$
lim_{g_{\mu\nu}\to\eta_{\mu\nu}}\, L = L_0 = \frac{1}{2}\frac{\sum\limits_{k=-\infty} b_k\,|\psi_k|exp[\frac{iS_{(k)}}{\hbar}]L_{(k)}}{\sum\limits_{k} b_k\,|\psi_k|\,exp[\frac{iS_{(k)}}{\hbar}]} + \frac{1}{2}\frac{\sum\limits_{k=-\infty} b^*_k\,|\psi_k|exp[\frac{-iS_{(k)}}{\hbar}]L_{(k)}}{\sum\limits_{k} b^*_k\,|\psi_k|\,exp[\frac{-iS_{(k)}}{\hbar}]}, \tag{2.2.13}
$$

$$
lim_{g_{\mu\nu}\to\eta_{\mu\nu}}\, L_{mix} = 0, \tag{2.2.14}
$$

$$
lim_{g_{\mu\nu}\to\eta_{\mu\nu}}\, T_{\mu}{}^{\nu} = T_{0\mu}{}^{\nu} = \frac{1}{2}\frac{\sum\limits_{k=-\infty} b_k\,|\psi_k|exp[\frac{iS_{(k)}}{\hbar}]T_{0(k)\mu}{}^{\nu}}{\sum\limits_{k} b_k\,|\psi_k|\,exp[\frac{iS_{(k)}}{\hbar}]} + \frac{1}{2}\frac{\sum\limits_{k=-\infty} b^*_k\,|\psi_k|exp[\frac{-iS_{(k)}}{\hbar}]T_{0(k)\mu}{}^{\nu}}{\sum\limits_{k} b^*_k\,|\psi_k|\,exp[\frac{-iS_{(k)}}{\hbar}]} \tag{2.2.15}
$$

where $lim_{g_{\mu\nu}\to\eta_{\mu\nu}}\, T_{(k)\mu}{}^{\nu} = T_{0(k)\mu}{}^{\nu} =$

$$
lim_{g_{\mu\nu}\to\eta_{\mu\nu}}\, T_{mix\mu}{}^{\nu} = 0\,. \tag{2.2.16}
$$

In the quantum case, due to the force generated by the quantum potential (e.g., responsible of the ballistic expansion of an isolated Gaussian packet) for the gradient of the EITD, it follows that

$$
T_{\mu}{}^{\nu}{}_{;\nu} = \left(|\psi_k|^2\, T_{\mu}{}^{\nu}\right)_{;\nu} \neq 0, \tag{2.2.17}
$$

Finally, it is worth noting that, since in the classical limit (whose resolution length is much bigger than the De Broglie length) due to the quantum decoherence [21] (produced by fluctuations) the superposition of states (2.1.1) undergo collapse to an eigenstate, the classical macroscopic limit is obtained by the limiting procedure

$$
lim_{macro} \equiv lim_{dec}\, lim_{\hbar\to 0} = lim_{\hbar\to 0}\, lim_{dec}\,. \tag{2.2.18}
$$

where the subscript "*dec*" stands for *decoherence* and where

$$
\begin{aligned}
lim_{dec}\,\psi &= lim_{dec} \sum_{k=k_{min}}^{k=k_{max}} b_k\,|\psi_k|\,exp[\frac{iS_{(k)}}{\hbar}] \\
&= b_{\tilde{k}}\,|\psi_{\tilde{k}}|\,exp[\frac{iS_{(\tilde{k})}}{\hbar}]
\end{aligned} \tag{2.2.19}
$$

where $k_{min} \le \tilde{k} \le k_{max}$. The exchange of the order of the two limits it is possible since, by (2.1.20), the $lim_{\hbar\to 0}$ applied to the eigenstates motion equation leads to the classical limit.

Since the detailed stochastic hydrodynamic derivation of (2.2.18) shows that the quantum non-local interactions can extend themselves beyond the De Broglie length in the case of strong coupling [21], here, we assume that (2.2.18) generally holds in a curved space-time taking into account the possibility that the macroscopic scale may go very much beyond the De Broglie length.

Finally, it is useful to note that in the Minkowskian case, by using (2.1.19,2.2.5,2.2.6), (2.2.8) reads

$$
\begin{aligned}
lim_{macro}\, \mathsf{T}_{0«}{}^{¬}{}_{;¬} &= lim_{\hbar\to 0}\, lim_{dec}\left( \mathsf{T}_{0Class«}{}^{¬}{}_{;¬} + \left( L_{0Class} ¢_{«}{}^{¬} \right)_{;¬} + \mathsf{T}_{0Q«}{}^{¬}{}_{;¬} + \mathsf{T}_{mix«}{}^{¬}{}_{;¬} \right) \\
&= lim_{\hbar\to 0}\left( \mathsf{T}_{(k)0Class«}{}^{¬}{}_{;¬} + \left( L_{(k)0Class} ¢_{«}{}^{¬} \right)_{;¬} + \mathsf{T}_{0(k)Q«}{}^{¬}{}_{;¬} \right) \qquad (2.2.20)\\
&= \mathsf{T}_{(k)0Class«}{}^{¬}{}_{;¬} + \left( L_{(k)0Class} ¢_{«}{}^{¬} \right)_{;¬} = 0
\end{aligned}
$$

**2.3 The minimum action in the hydrodynamic formalism**

Since the hydrodynamic Lagrangean depends also by the quantum potential and hence by $|\cdot| = f_1(|\cdot_k|)$ and $\partial_« |\cdot| = f_2(|\cdot_k|, \partial_« |\cdot_k|)$, the problem of defining the equation of motion can be generally carried out by using the set of variables : $x_{(k)«} = (q_«, \dot{q}_{«(k)}, |\cdot_k|, \partial_« |\cdot_k|)$. Thence, the variation of the hydrodynamic action $S = \int\iiint L\, dVdt$ (between the fixed starting and end points, $q_{«a}$ $q_{«b}$, respectively) where $L \equiv |\cdot|^2 L$, reads [19]

$$
\begin{aligned}
¢S &= \int\iiint |\cdot|^2 \left( \sum_k \frac{\partial \tilde{L}_{(k)}}{\partial x_{(k)}{}^{«}} \right) ¢\, x_{(k)}{}^{«}\, dVdt \\
&= \int\iiint |\cdot|^2 \sum_k \left( \frac{\partial \tilde{L}_{(k)}}{\partial q^{«}} ¢\, q^{«} + \frac{\partial \tilde{L}_{(k)}}{\partial \dot{q}^{«}_{(k)}} ¢\, \dot{q}^{«}_{(k)} + \frac{\partial \tilde{L}_{(k)}}{\partial |\cdot_k|} ¢\, |\cdot_k| + \frac{\partial \tilde{L}_{(k)}}{\partial \partial^{«} |\cdot_k|} ¢\partial^{«} |\cdot_k| \right) dVdt. \qquad (2.3.1)\\
&= \frac{1}{c}\int\iiint |\cdot|^2 \sum_k \left( \left( \frac{\partial \tilde{L}_{(k)}}{\partial q^{«}} - \frac{d}{dt}\frac{\partial \tilde{L}_{(k)}}{\partial \dot{q}^{«}_{(k)}} \right) ¢\, q^{«} + \left( \frac{\partial \tilde{L}_{(k)}}{\partial |\cdot_k|} - \partial^{«} \frac{\partial \tilde{L}_{(k)}}{\partial \partial^{«} |\cdot_k|} \right) ¢\, |\cdot_k| \right) d\Omega
\end{aligned}
$$

Given that the quantum motion equations for eigenstates (2.1.14-15) satisfy the condition

$$
\left( \frac{\partial \tilde{L}_{(k)}}{\partial q^{«}} - \frac{d}{dt}\frac{\partial \tilde{L}_{(k)}}{\partial \dot{q}^{«}_{(k)}} \right) = \left( \frac{\partial L_{(k)}}{\partial q^{«}} - \frac{d}{dt}\frac{\partial L_{(k)}}{\partial \dot{q}^{«}_{(k)}} \right) = 0 \ , \qquad (2.3.2)
$$

(that explicitly defines $\cdot_{(q,t)}$) the variation of the action $¢S$ for the k-th eigenstates reads

$$
¢S = ¢\left( \Delta S_{Q(k)} \right) = \frac{1}{c}\int\iiint |\cdot_k|^2 \left( \frac{\partial L_{(k)}}{\partial |\cdot_k|} - \partial^{«} \frac{\partial L_{(k)}}{\partial \partial_{«} |\cdot_k|} \right) ¢\, |\cdot_k|\, d\Omega \qquad (2.3.3)
$$

that is not null since it takes contribution from the quantum potential contained into the hydrodynamic Lagrangean $L_{(k)}$.

Thence, for the quantum hydrodynamic evolution it follows that

$$
¢S - ¢\left( \Delta S_{Q(k)} \right) = 0 \qquad (2.3.4)
$$

that, since in the classical limit, for $\hbar \to 0$, $V_{qu} \to 0$, it holds that

$$
\frac{\partial\left( lim_{\hbar\to 0} L_{(k)} \right)}{\partial |\cdot_k|} = 0, \qquad (2.3.5)
$$

$$\frac{\partial\left(lim_{\hbar\to 0} L_{(k)}\right)}{\partial^{\mu}|\psi_{k}|}=0, \qquad (2.3.6)$$

the classical extremal principle

$$lim_{\hbar\to 0}\delta S = lim_{\hbar\to 0}\delta\left(\Delta S_{Q(k)}\right)=0 \ . \qquad (2.3.7)$$

is recovered.

Moreover, generally speaking, by using (2.1.21), for the general superposition of state (2.1.1) the variation of the action reads

$$\begin{aligned}\delta S &= -\frac{1}{c}\int\iiint|\psi|^{2}\sum_{k}\left(\left(\tilde{p}_{(k)}\frac{\partial\dot{q}_{(k)\nu}}{\partial q^{\mu}}\right)\delta q^{\mu}-\left(\frac{\partial\tilde{L}_{(k)}}{\partial|\psi_{k}|}-\partial^{\mu}\frac{\partial\tilde{L}_{(k)}}{\partial\partial^{\mu}|\psi_{k}|}\right)\delta|\psi_{k}|\right)d\Omega \\ &= \delta\left(\Delta S_{Q_{mix}}+\Delta S_{Q}\right)=\delta\left(\Delta S\right)\end{aligned} \qquad (2.3.8)$$

where

$$\Delta S_{Q}=\frac{1}{c}\int\iiint|\psi|^{2}\sum_{k}\left(\frac{\partial\tilde{L}_{(k)}}{\partial|\psi_{k}|}-\partial^{\mu}\frac{\partial\tilde{L}_{(k)}}{\partial\partial^{\mu}|\psi_{k}|}\right)\delta|\psi_{k}|d\Omega \qquad (2.3.9)$$

and where

$$\delta\left(\Delta S_{Q_{mix}}\right)=-\frac{1}{c}\int\iiint|\psi|^{2}\sum_{k}\tilde{p}_{(k)}\frac{\partial\dot{q}_{(k)\nu}}{\partial q^{\mu}}\delta|\psi_{k}|d\Omega \qquad (2.3.10)$$

is due to the quantum mixing of superposition of states.
Thence, by using (2.2.3), the condition (2.3.8) can be generalized to

$$\delta S-\delta\left(\Delta S\right)=0 \qquad (2.3.11)$$

that, being

$$lim_{macro}\,\delta\left(\Delta S_{Qmix}\right)=lim_{\hbar\to 0}\,lim_{dec}\,\delta\left(\Delta S_{Qmix}\right)=0, \qquad (2.3.12)$$

in the classical limit leads to

$$\begin{aligned}lim_{macro}\,\delta S &= lim_{macro}\,\delta\left(\Delta S\right)=lim_{\hbar\to 0}\,lim_{dec}\,\delta\left(\Delta S_{Q}+\Delta S_{Qmix}\right) \\ &= lim_{\hbar\to 0}\,\delta\left(\Delta S_{Q(k)}\right)=0\end{aligned} \ . \qquad (2.3.13)$$

**2.4. The hydrodynamic KGE in the curvilinear space-time**

As far as it concerns the motion equations (2.1.8-9), there is no way to univocally define them in the non-Minkowskian space-time without a postulate that fixes the criterion of generalization.

In the classical general relativity this criterion is given by the equivalence of inertial and gravitational mass that, by fact, is equivalent to postulate that the classical equation of motion is covariant in general relativity [19] .

Analogously, assuming the covariance of the KGE [19] that reads

$$\psi_{;\mu}^{;\mu}=\left(g^{\mu\nu}\partial_{\nu}\psi\right)_{;\mu}=\frac{1}{\sqrt{-g}}\partial_{\mu}\sqrt{-g}\left(g^{\mu\nu}\partial_{\nu}\psi\right)=-\frac{m^{2}c^{2}}{\hbar^{2}}\psi \ . \qquad (2.4.1)$$

also the hydrodynamic motion equations (2.0.2-3) own a covariant form [19] and they read

$$g_{«¬}\partial^{¬} S\ \partial^{«} S-\hbar^{2}\frac{1}{|\cdot|\sqrt{-g}}\partial_{«}\sqrt{-g}\left(g^{«¬}\partial_{¬}|\cdot|\right)-m^2c^2=0 \tag{2.4.2}$$

$$\frac{1}{\sqrt{-g}}\frac{\partial}{\partial q^{«}}\sqrt{-g}\left(g^{«¬}|\cdot|^{2}\frac{\partial S}{\partial q^{¬}}\right)=0 \tag{2.4.3}$$

where the quantum potential reads

$$V_{qu}=-\frac{\hbar^2}{m}\frac{1}{|\cdot|\sqrt{-g}}\partial_{«}\sqrt{-g}\left(g^{«¬}\partial_{¬}|\cdot|\right), \tag{2.4.4}$$

where $g_{¬«}$ is the metric tensor and where $g=|g_{¬«}|^{-1}$.

Moreover, given the covariance of (2.4.2-3) also the motion equation (2.1.20 or 2.1.21) as well as $\mathsf{T}_{(k)«¬}$ and $L_{(k)}$, that read respectively

$$\begin{aligned}\mathsf{T}_{(k)«¬}&=\left(\dot{q}_{¬}\frac{\partial L_{(k)}}{\partial\dot{q}^{«}}-L_{(k)}¢_{«}^{¬}\right)\\&=mc^2\sqrt{\frac{g_{©ª}\dot{q}^{ª}_{(k)}\dot{q}^{©}_{(k)}}{c^2}}\sqrt{1-\frac{V_{qu(k)}}{mc^2}}\left(u_{«}u_{¬}-g_{~}u^{~}u\ g_{«¬}\right)\end{aligned}, \tag{2.4.5}$$

where

$$\begin{aligned}L_{(k)}&=-\frac{mc^2}{¥_{(k)}}\sqrt{1-\frac{V_{qu(k)}}{mc^2}}=-mc^2\sqrt{\frac{g_{«¬}\dot{q}^{¬}_{(k)}\dot{q}^{«}_{(k)}}{c^2}}\sqrt{1-\frac{V_{qu(k)}}{mc^2}},\\&=-g_{«~}\dot{q}^{~}_{(k)}p^{«}{}_{(k)}=-g_{«~}c^2\left(\partial_t S_{(k)}\right)^{-1}p^{~}{}_{(k)}p^{«}{}_{(k)}\end{aligned} \tag{2.4.6}$$

are covariant.

Once the quantum equations are defined in the non-Minkowskian space-time, their meaning is fully determined when the metric of the space-time is defined by the gravity equation (GE) based on additional condition (e.g., on the hydrodynamic action covariantly generalized).

Moreover, it is useful to note that, due to the biunique relation between the quantum hydrodynamic equations (2.4.2-3), subject to the irrotational conditions, and the KGE (2.4.1) [17], the system of equations (2.4.2-3) (as well as (2.1.20 or 2.1.21) submitted to current conservation and irrotational conditions) coupled to the GE, are equivalent to the KGE-GE system where both the EIT (2.2.7) and the Lagrangean (2.1.5) can be expressed as a function of the field $\cdot$ by using the following relations (2.0.12, 2.0.13, 2.2.1),

$$\begin{aligned}L_{(k)}&=-c^2\left(\partial_t S_{(k)}\right)^{-1}g_{«~}p^{~}{}_{(k)}p^{«}{}_{(k)}=-c^2\left(\partial_t S_{(k)}\right)^{-1}g_{«~}\partial^{~}S_{(k)}\partial^{«}S_{(k)}\\&=-c^2\frac{i\hbar}{2}\left(\frac{\partial\, ln[\frac{\cdot_k}{\cdot_k{}^*}]}{\partial t}\right)^{-1}g_{«~}\frac{\partial\, ln[\frac{\cdot_k}{\cdot_k{}^*}]}{\partial q_{~}}\frac{\partial\, ln[\frac{\cdot_k}{\cdot_k{}^*}]}{\partial q_{¬}}\end{aligned} \tag{2.4.7}$$

$$\mathrm{T}_{(k)«¬} = c^2\left(\frac{\partial S_{(k)}}{\partial t}\right)^{-1}\left(\frac{\partial S_{(k)}}{\partial q^«}\frac{\partial S_{(k)}}{\partial q^¬} - g_{\sim}\ \frac{\partial S_{(k)}}{\partial q}\frac{\partial S_{(k)}}{\partial q_{\sim}} g_{«¬}\right)$$

$$= -m^2c^4\left(\frac{\frac{\hbar}{2i}\partial\, ln[\frac{·_k}{·_k{}^*}]}{\partial t}\right)^{-1}\left(\left(\frac{\hbar}{2mc}\right)^2\frac{\partial\, ln[\frac{·_k}{·_k{}^*}]}{\partial q^«}\frac{\partial\, ln[\frac{·_k}{·_k{}^*}]}{\partial q^¬} + \left(1-\frac{V_{qu(k)}}{mc^2}\right)g_{«¬}\right) \quad (2.4.8)$$

and

$$\dot{q}_{(k)«} = c^2\frac{p_{(k)«}}{E} = -c^2\frac{\partial_« S_{(k)}}{\partial_t S_{(k)}} = -c^2\frac{\partial_«\, ln[\frac{·_k}{·_k{}^*}]}{\partial_t\, ln[\frac{·_k}{·_k{}^*}]} \quad (2.4.9)$$

where the last identity has been obtained by inverting (2.0.13).
It is worth mentioning that the KGE-GE system of evolutionary equations owns the advantage to contain only the irrotational states that satisfy the current conservation condition.

## 3. The minimum action in curved space-time and the gravity equation for the hydrodynamic KGE

In order to follow an analytical procedure we derive the gravity equation, by applying the minimum action principle to the quantum hydrodynamic matter evolution associated to the field $·$ .

Given that the quantum hydrodynamic equations in the Minkowskian space-time [19], satisfies the minimum action principle (2.3.11), when we consider the covariant formulation in the curved space-time, such variation takes a contribution from the variability of the metric tensor. When we consider the gravity and we assume that the geometry of the space-time is that one which makes null the overall variation of the action (2.3.11), we define the condition that leads to the definition of the GE.

By considering the variation of the action due to the curvilinear coordinates [22] and the functional dependence by $·$ , it follows that

$$¢S = \frac{1}{c}\int\iiint |·|^2 \sum_k \left(\begin{array}{l}\left(\frac{1}{\sqrt{-g}}\left(\frac{\partial\sqrt{-g}\tilde{L}_{(k)}}{\partial g^{«¬}} - \frac{\partial}{\partial q^ª}\frac{\partial\sqrt{-g}\tilde{L}_{(k)}}{\partial\frac{\partial g^{«¬}}{\partial q^ª}} - \frac{\partial}{\partial|·|}\frac{\partial\sqrt{-g}\tilde{L}_{(k)}}{\partial\frac{\partial g^{«¬}}{\partial|·_k|}}\right)\right)¢g^{«¬} \\ +\tilde{p}_{(k)¬}\dot{q}_{(k)}{}^{¬}{}_{;«}¢q^« + \left(\frac{\partial\tilde{L}_{(k)}}{\partial|·_k|} - \partial^«\frac{\partial\tilde{L}_{(k)}}{\partial\partial^«|·_k|}\right)¢|·_k|\end{array}\right)\sqrt{-g}\,d\Omega \quad (3.1)$$

and

$$¢S - ¢(\Delta S) = \frac{1}{c}\int\iiint |·|^2 \sum_k\left(\frac{\partial\sqrt{-g}\tilde{L}_{(k)}}{\partial g^{«¬}} - \frac{\partial}{\partial q^ª}\frac{\partial\sqrt{-g}\tilde{L}_{(k)}}{\partial\frac{\partial g^{«¬}}{\partial q^ª}} - \frac{\partial}{\partial|·_k|}\frac{\partial\sqrt{-g}\tilde{L}_{(k)}}{\partial\frac{\partial g^{«¬}}{\partial|·_k|}}\right)¢g^{«¬}d\Omega\,. \quad (3.2)$$

If we postulate that the variation of action of the gravitational field

$$¢S_g = \frac{c^3}{16®G}\int\iiint\left(R_{«¬} - \frac{1}{2}Rg_{«¬}\right)¢g^{«¬}\sqrt{-g}\,d\Omega \tag{3.3}$$

offsets that one produced by the KGE field so that

$$¢S - ¢(\Delta S) + ¢S_g = \int\iiint\left(R_{«¬} - \frac{1}{2}Rg_{«¬} - \frac{8®G}{c^4}|·|^2\,²_{«¬}\right)¢g^{«¬}\sqrt{-g}\,d\Omega = 0, \tag{3.4}$$

we obtain the gravitational equation

$$R_{«¬} - \frac{1}{2}Rg_{«¬} = \frac{8®G}{c^4}|·|^2\,²_{«¬} \tag{3.5}$$

where

$$²_{«¬} = \frac{1}{2}\frac{\sum_{k=-\infty} b_k |·_k| exp[\frac{iS_{(k)}}{\hbar}]\left(²_{(k)«¬class} + ²_{(k)«¬Q} + ²_{(k)«¬mix}\right)}{\sum_k b_k |·_k| exp[\frac{iS_{(k)}}{\hbar}]}$$

$$+\frac{1}{2}\frac{\sum_{k=-\infty} b*_k |·_k| exp[\frac{-iS_{(k)}}{\hbar}]\left(²_{(k)«¬class} + ²_{(k)«¬Q} - ²_{(k)«¬mix}\right)}{\sum_k b*_k |·_k| exp[\frac{-iS_{(k)}}{\hbar}]} \tag{3.6}$$

$$+ ²_{«¬curv}$$

$$= ²_{«¬class} + ²_{«¬Q} + ²_{«¬mix} + ²_{«¬curv}$$

where

$$²_{(k)«¬class} = 2\frac{1}{\sqrt{-g}}\left(\frac{\partial}{\partial g^{«¬}} - \frac{\partial}{\partial q^{ª}}\frac{\partial}{\partial\frac{\partial g^{«¬}}{\partial q^{ª}}} - \frac{\partial}{\partial|·_k|}\frac{\partial}{\partial\frac{\partial g^{«¬}}{\partial|·_k|}}\right)\sqrt{-g}L_{(k)class} \tag{3.7}$$

$$²_{(k)«¬Q} = 2\frac{1}{\sqrt{-g}}\left(\frac{\partial}{\partial g^{«¬}} - \frac{\partial}{\partial q^{ª}}\frac{\partial}{\partial\frac{\partial g^{«¬}}{\partial q^{ª}}} - \frac{\partial}{\partial|·_k|}\frac{\partial}{\partial\frac{\partial g^{«¬}}{\partial|·_k|}}\right)\sqrt{-g}L_{(k)Q} \tag{3.8}$$

$$²_{(k)«¬mix} = 2\frac{1}{\sqrt{-g}}\left(\frac{\partial}{\partial g^{«¬}} - \frac{\partial}{\partial q^{ª}}\frac{\partial}{\partial\frac{\partial g^{«¬}}{\partial q^{ª}}} - \frac{\partial}{\partial|·_k|}\frac{\partial}{\partial\frac{\partial g^{«¬}}{\partial|·_k|}}\right)\sqrt{-g}L_{(k)mix} \tag{3.9}$$

$$
{}^{2}{}_{\pm i\text{«¬}curv}=\frac{1}{2}\sum_{k=-\infty}\begin{pmatrix}L_{(k)\text{«¬}class}\\+L_{(k)\text{«¬}Q}\\+L_{(k)\text{«¬}mix}\end{pmatrix}\left(\begin{pmatrix}\frac{\partial}{\partial g^{\text{«¬}}}-\frac{\partial}{\partial q^{a}}\frac{\partial}{\partial\frac{\partial g^{\text{«¬}}}{\partial q^{a}}}\\-\frac{\partial}{\partial|\cdot_{\pm i(k)}|}\frac{\partial}{\partial\frac{\partial g^{\text{«¬}}}{\partial|\cdot_{\pm i(k)}|}}\end{pmatrix}\Xi_{(k)}-\begin{pmatrix}\frac{\partial\Xi_{(k)}}{\partial q^{a}}\frac{\partial ln\left(L_{(k)class}\right)}{\frac{\partial g^{\text{«¬}}}{\partial q^{a}}}+\frac{\partial ln\left(L_{(k)class}\right)}{\partial q^{a}}\frac{\partial\Xi_{(k)}}{\frac{\partial g^{\text{«¬}}}{\partial q^{a}}}\\\frac{\partial\Xi_{(k)}}{\partial|\cdot_{k}|}\frac{\partial ln\left(L_{(k)class}\right)}{\frac{\partial g^{\text{«¬}}}{\partial|\cdot_{k}|}}+\frac{\partial ln\left(L_{(k)class}\right)}{\partial|\cdot_{k}|}\frac{\partial\Xi_{(k)}}{\frac{\partial g^{\text{«¬}}}{\partial|\cdot_{k}|}}\end{pmatrix}\right)
$$

$$
+\frac{1}{2}\sum_{k=-\infty}\begin{pmatrix}L_{(k)\text{«¬}class}\\+L_{(k)\text{«¬}Q}\\-L_{(k)\text{«¬}mix}\end{pmatrix}\left(\begin{pmatrix}\frac{\partial}{\partial g^{\text{«¬}}}-\frac{\partial}{\partial q^{a}}\frac{\partial}{\partial\frac{\partial g^{\text{«¬}}}{\partial q^{a}}}\\-\frac{\partial}{\partial|\cdot_{\pm i(k)}|}\frac{\partial}{\partial\frac{\partial g^{\text{«¬}}}{\partial|\cdot_{\pm i(k)}|}}\end{pmatrix}\Xi^{*}_{(k)}-\begin{pmatrix}\frac{\partial\Xi^{*}_{(k)}}{\partial q^{a}}\frac{\partial ln\left(L_{(k)class}\right)}{\frac{\partial g^{\text{«¬}}}{\partial q^{a}}}+\frac{\partial ln\left(L_{(k)class}\right)}{\partial q^{a}}\frac{\partial\Xi^{*}_{(k)}}{\frac{\partial g^{\text{«¬}}}{\partial q^{a}}}\\\frac{\partial\Xi^{*}_{(k)}}{\partial|\cdot_{k}|}\frac{\partial ln\left(L_{(k)class}\right)}{\frac{\partial g^{\text{«¬}}}{\partial|\cdot_{k}|}}+\frac{\partial ln\left(L_{(k)class}\right)}{\partial|\cdot_{k}|}\frac{\partial\Xi^{*}_{(k)}}{\frac{\partial g^{\text{«¬}}}{\partial|\cdot_{k}|}}\end{pmatrix}\right) \quad (3.10)
$$

where

$$
\Xi_{(k)}=\frac{b_{k}|\cdot_{k}|exp[\frac{iS_{(k)}}{\hbar}]}{\sum_{j}b_{j}|\cdot_{j}|exp[\frac{iS_{(k)}}{\hbar}]}.
$$

Moreover, given that [22]

$$
\frac{1}{2c}\int\iiint{}^{2}{}_{\text{«¬}}\text{¢}g^{\text{«¬}}\sqrt{-g}d\Omega=-\frac{1}{c}\int\iiint{}^{2\text{¬}}{}_{\text{«;¬}}\text{¶}^{\text{«}}\sqrt{-g}d\Omega \quad (3.11)
$$

where

$$
\text{¢}g^{\text{«¬}}=\text{¶}^{\text{«;¬}}+\text{¶}^{\text{¬;«}}, \quad (3.12)
$$

for the infinitesimal transformation of coordinates

$$q'^{«} = q^{«} + ¶^{«} , \qquad (3.13)$$

around the Minkowskian case (3.11) leads to

$$¢S - ¢(\Delta S) = -\frac{1}{c}\int\iiint \left( |\cdot|^{2}\, 2^{¬}_{«} \right)_{;¬} ¶^{«} \sqrt{-g}\, d\Omega = 0 \qquad (3.14)$$

and, in the classical limit, to

$$\begin{aligned} &lim_{decoherence}\, lim_{\hbar \to 0} -\frac{1}{c}\int\iiint \left( |\cdot|^{2}\, 2^{¬}_{«} \right)_{;¬} ¶^{«} \sqrt{-g}\, d\Omega = \\ &-\frac{1}{c}\int\iiint \left( lim_{decoherence}\, lim_{\hbar \to 0} \left( |\cdot|^{2}\, 2^{¬}_{«} \right)_{;¬} \right) ¶^{«} \sqrt{-g}\, d\Omega = 0 \end{aligned} \qquad (3.15)$$

and thence, for the arbitrariness of $¶^{«}$, to

$$lim_{dec}\, lim_{\hbar \to 0} \left( |\cdot|^{2}\, 2^{¬}_{«} \right)_{;¬} = 0 . \qquad (3.16)$$

Furthermore, given that from (2.2.6, 2.2.8-11) in the classical (Minkowskian) limit it holds

$$\begin{aligned} &lim_{dec}\, lim_{\hbar \to 0} \left( |\cdot|^{2}\, \mathrm{T}^{¬}_{«} \right)_{;¬} = \left( lim_{dec}\, lim_{\hbar \to 0} |\cdot|^{2}\, \mathrm{T}^{¬}_{«} \right)_{;¬} \\ &= \left( lim_{dec} |\cdot|^{2} \left( \mathrm{T}^{¬}_{«\, class} + L_{class} ¢^{¬}_{«} + \mathrm{T}^{¬}_{mix\, «} \right) \right)_{;¬} = |\cdot_{k}|^{2}\, \mathrm{T}^{¬}_{(k)«;¬\, class} + |\cdot_{k}|^{2} \left( L_{(k) class} ¢^{¬}_{«} \right)_{;¬} = 0 \end{aligned} , \qquad (3.17)$$

from (3.6) it follows that

$$\begin{aligned} lim_{dec}\, lim_{\hbar \to 0} |\cdot|^{2}\, 2_{«}^{¬} &= |\cdot_{k}|^{2}\, 2_{(k)«}^{¬}{}_{class} \\ &= |\cdot_{k}|^{2}\, \mathrm{T}_{(k)«}^{¬}{}_{class} + |\cdot_{k}|^{2}\, L_{(k) class} ¢^{¬}_{«} + |\cdot_{k}|^{2}\, C_{(\dot{q}_{«})} ¢_{«}^{¬} \end{aligned} \qquad (3.18)$$

and, that

$$lim_{dec}\, lim_{\hbar \to 0} \left( R_{«¬} - \frac{1}{2} R g_{«¬} \right) = -\frac{8®G}{c^{4}} \left( T_{(k)«¬ class} + |\cdot|^{2}\, L_{(k) class}\, g_{«¬} + |\cdot|^{2}\, C g_{«¬} \right) . \qquad (3.19)$$

where

$$T_{(k)«¬ class} = |\cdot_{k}|^{2}\, \mathrm{T}_{(k)«¬ class} . \qquad (3.20)$$

Moreover, by posing

$$\Lambda_{class} = -L_{(k) class (\dot{q}_{«})} - C_{(\dot{q}_{«})} , \qquad (3.22)$$

the Einstein equation (where the k variable is needless)

$$lim_{decoherence}\, lim_{\hbar \to 0} \left( R_{«¬} - \frac{1}{2} R g_{«¬} \right) = R^{macro}_{«¬} - \frac{1}{2} R^{macro} g_{«¬} = \frac{8®G}{c^{4}} \left( T_{(k)«¬ class} - \Lambda g_{«¬} \right) , \qquad (3.23)$$

where

$$\Lambda = |\cdot_{k}|^{2}\, \Lambda_{class} , \qquad (3.24)$$

is recovered in the classical limit

### 3.1. The GE for the KGE eigenstates

Moreover, by using the identity

$$^{2}{}_{(k)\mu\nu\,class} = T_{(k)\mu\nu\,class} - \Lambda_{class} ¢_{\mu\nu} , \qquad (3.1.1)$$

obtained from (3.18, 3.22) and by using (2.1.18, 3.8) it follows that

$$\begin{aligned} ^{2}{}_{(k)\mu\nu Q} &= -\sim_{(V_{qu})}\left( ^{2}{}_{(k)\mu\nu class} + L_{(k)class}\Delta_{\mu\nu(k)} \right) \\ &= -\sim_{(V_{qu})}\left( T_{(k)\mu\nu class} - \Lambda_{class} g_{\mu\nu} + L_{(k)class}\Delta_{\mu\nu(k)} \right) \end{aligned} \qquad (3.1.2)$$

where

$$\Delta_{\mu\nu(k)} = \frac{2}{\sim_{(V_{qu(k)})}} \left( \begin{array}{c} \left( \frac{\partial}{\partial g^{\mu\nu}} - \frac{\partial}{\partial q^{\alpha}} \frac{\partial}{\partial \frac{\partial g^{\mu\nu}}{\partial q^{\alpha}}} - \frac{\partial}{\partial |\cdot_k|} \frac{\partial}{\partial \frac{\partial g^{\mu\nu}}{\partial |\cdot_k|}} \right) \sim_{(V_{qu(k)})} \\ - \left( \begin{array}{c} \frac{\partial \sim_{(V_{qu(k)})}}{\partial q^{\alpha}} \frac{\partial ln\left(L_{(k)class}\right)}{\partial \frac{\partial g^{\mu\nu}}{\partial q^{\alpha}}} + \frac{\partial ln\left(L_{(k)class}\right)}{\partial q^{\alpha}} \frac{\partial \sim_{(V_{qu(k)})}}{\partial \frac{\partial g^{\mu\nu}}{\partial q^{\alpha}}} \\ \frac{\partial \sim_{(V_{qu(k)})}}{\partial |\cdot_k|} \frac{\partial ln\left(L_{(k)class}\right)}{\partial \frac{\partial g^{\mu\nu}}{\partial |\cdot_k|}} + \frac{\partial ln\left(L_{(k)class}\right)}{\partial |\cdot_k|} \frac{\partial \sim_{(V_{qu(k)})}}{\partial \frac{\partial g^{\mu\nu}}{\partial |\cdot_k|}} \end{array} \right) \end{array} \right) , \qquad (3.1.3)$$

and, being both $^{2}{}_{\mu\nu curv} = 0$ and $^{2}{}_{\mu\nu mix} = 0$, that

$$R_{\mu\nu} - \frac{1}{2} R g_{\mu\nu} + \frac{8\pi G}{c^4} \Lambda g_{\mu\nu} = \frac{8\pi G}{c^4} \left( \begin{array}{c} T_{(k)\mu\nu class}\left(1 - \sim_{(V_{qu(k)})}\right) \\ + \sim_{(V_{qu(k)})}\left( \Lambda g_{\mu\nu} - |\cdot_k|^2 L_{(k)class}\Delta_{\mu\nu(k)} \right) \end{array} \right) . \qquad (3.1.4)$$

Finally, by separating $\Delta_{\mu\nu}$ in the isotropic and stress part $\Delta_{s\mu\nu}$ as following

$\Delta_{\mu\nu} = \frac{\Delta_{\alpha\alpha}}{4} g_{\mu\nu} + \Delta_{s\mu\nu}$ , the GE, for eigenstates, reads

$$R_{\mu\nu} - \frac{1}{2} R g_{\mu\nu} + \frac{8\pi G}{c^4} \Lambda g_{\mu\nu} = \frac{8\pi G}{c^4} \left( \begin{array}{c} T_{(k)\mu\nu class}\left(1 - \sim_{(V_{qu(k)})}\right) \\ + \sim_{(V_{qu(k)})} \left( \begin{array}{c} \left( \Lambda - |\cdot_k|^2 L_{(k)class} \frac{\Delta_{\alpha\alpha}}{4} \right) g_{\mu\nu} \\ - |\cdot_k|^2 L_{(k)class}\Delta_{s\mu\nu(k)} \end{array} \right) \end{array} \right) \qquad (3.1.5)$$

### 3.2. The GE of the general KGE field

Finally, by using (3.6-3.10) the GE as a function of the general KGE field (2.1.1) reads

$$R_{\text{«¬}} - \frac{1}{2} R g_{\text{«¬}} + \frac{8\text{®}G}{c^4}\Lambda g_{\text{«¬}} = \frac{8\text{®}G}{c^4} |\cdot\ |^2\, Tr\left( \text{²}_{\text{«¬}\,class} - \text{i}_{Q}\, g_{\text{«¬}} + \text{a}\,\text{²}_{\text{«¬}\,stress}\right) \quad (3.2.1)$$

where

$$\text{²}_{\text{«¬}\,class} = \text{s}_{(k)}\left(1 - \sim_{(V_{qu(k)})}\right) T_{(k)\text{«¬}\,class}\,\text{¢}_{hk}\,, \quad (3.2.2)$$

where

$$\text{s}_{(k)} = \left( \begin{array}{l} \dfrac{1}{2}\dfrac{b_k |\cdot_k| exp[\frac{iS_{(k)}}{\hbar}]}{\sum_j b_j |\cdot_j| exp[\frac{iS_{(j)}}{\hbar}]} \\ + \dfrac{1}{2}\dfrac{b*_k |\cdot_k| exp[\frac{-iS_{(k)}}{\hbar}]}{\sum_j b*_j |\cdot_j| exp[\frac{-iS_{(j)}}{\hbar}]} \end{array} \right) \quad (3.2.3)$$

$$\text{i}_{Q} = -\left( \left( \begin{array}{l} \dfrac{1}{2}\dfrac{b_k |\cdot_k| exp[\frac{iS_{(k)}}{\hbar}] \left( \begin{array}{l} \sim_{(V_{qu(k)})}\left( \Lambda_{class} - L_{(k)class}\dfrac{\Delta_{\text{©©}(k)}}{4} \right) \\ + \dfrac{\text{²}_{(k)\text{©©}\,mix}}{4} \end{array} \right)}{\sum_j b_j |\cdot_j| exp[\frac{iS_{(j)}}{\hbar}]} \\ + \dfrac{1}{2}\dfrac{b*_k |\cdot_k| exp[\frac{-iS_{(k)}}{\hbar}] \left( \begin{array}{l} \sim_{(V_{qu(k)})}\left( \Lambda_{class} - L_{(k)class}\dfrac{\Delta_{\text{©©}(k)}}{4} \right) \\ - \dfrac{\text{²}_{(k)\text{©©}\,mix}}{4} \end{array} \right)}{\sum_j b*_j |\cdot_j| exp[\frac{-iS_{(j)}}{\hbar}]} \end{array} \right) + \frac{\text{²}_{\text{aa}\,curv}}{4} \right) \text{¢}_{hk}\,, \quad (3.2.4)$$

and

$$
\text{a}^2{}_{\text{«¬}stress} = \left( \left( \begin{array}{c} \frac{1}{2}\frac{b_k\,|\cdot{}_k|\,exp[\frac{iS_{(k)}}{\hbar}]\left(\begin{array}{c}\left(\sim_{(V_{qu(k)})}L_{(k)class}\Delta_{s\text{«¬}(k)}\right)\\ -^2{}_{s(k)\text{«¬}mix}\end{array}\right)}{\sum_j b_j\,|\cdot{}_j|\,exp[\frac{iS_{(j)}}{\hbar}]} \\ +\frac{1}{2}\frac{b^*{}_k\,|\cdot{}_k|\,exp[\frac{-iS_{(k)}}{\hbar}]\left(\begin{array}{c}\left(\sim_{(V_{qu(k)})}L_{(k)class}\Delta_{s\text{«¬}(k)}\right)\\ +^2{}_{s(k)\text{«¬}mix}\end{array}\right)}{\sum_j b^*{}_j\,|\cdot{}_j|\,exp[\frac{-iS_{(j)}}{\hbar}]} \end{array} \right) +{}^2{}_{s\text{«¬}curv} \right) \text{¢}_{hk} \quad (3.2.5)
$$

where both ${}^2{}_{(k)\text{«¬}mix}$ and ${}^2{}_{\text{«¬}curv}$ have been splitted into the isotropic and stress parts as in the following

$$
{}^2{}_{(k)\text{«¬}mix} = \frac{{}^2{}_{(k)\text{aa}\,mix}}{4} g_{\text{«¬}} +{}^2{}_{s(k)\text{«¬}mix} \quad (3.2.6)
$$

$$
{}^2{}_{\text{«¬}curv} = \frac{{}^2{}_{\text{aa}\,curv}}{4} g_{\text{«¬}} +{}^2{}_{s\text{«¬}curv} . \quad (3.2.7)
$$

Equation (3.1.4-5, 3.2.1) can be expressed as a function of the KGE field by using the relations (2.2.4, 2.4.7-2.4.9). Actually, the hydrodynamic approach has been used as a "Trojan horse" to find the GE where the non-physical states are implicitly excluded by writing it as a function of the KGE field (i.e., we do not need to impose the irrotational condition).

The main difference with the Einstein equation is given by the terms $\sim_{(V_{qu})}$, $\text{a}^2{}_{\text{«¬}stress}$, ${}^2{}_{\text{«¬}mix}$ and ${}^2{}_{\text{«¬}curv}$ whose quantum-mechanical origin can be noticed by passing to the macroscopic classical scale being $lim_{\hbar\to 0}\sim_{(V_{qu})} = 0$ , $lim_{macro}\text{a}^2{}_{\text{«¬}stress} = 0$, $lim_{macro}{}^2{}_{\text{«¬}curv} = 0$ and $lim_{macro}{}^2{}_{\text{«¬}mix} = 0$.

Finally, it must be noted that equation (3.1.5) represents the decoherent limit of (3.2.1).

## 4. Perturbative approach to the GE- KGE system

For particles very far from the Planckian mass density $\frac{m_p}{l_p{}^3} = \frac{c^5}{\hbar G^2}$, it is possible to solve the system of equation

$$
R_{\text{«¬}} - \frac{1}{2}Rg_{\text{«¬}} + \frac{8\text{®}G}{c^4}\Lambda g_{\text{«¬}} = \frac{8\text{®}G}{c^4}|\cdot|^2\, Tr\left({}^2{}_{\text{«¬}class(\cdot)} - \text{i}_{\,Q}\, g_{\text{«¬}(\cdot)} + \text{a}^2{}_{\text{«¬}stress(\cdot)}\right) \quad (4.1)
$$

$$
\left(g^{\text{«¬}}\partial_{\text{¬}}\cdot\right)_{;\text{«}} = -\frac{m^2c^2}{\hbar^2}\cdot \quad (4.2)
$$

by a perturbative iteration

$$
R_{\text{«¬}\left(\text{£}_{\text{«¬}}\right)} \cong R^{(0)}{}_{\text{«¬}\left(\text{£}_{\text{«¬}}\right)} + R^{(1)}{}_{\text{«¬}\left(\text{£}_{\text{«¬}}\right)} + R^{(2)}{}_{\text{«¬}\left(\text{£}_{\text{«¬}}\right)} + \dots , \quad (4.3)
$$

$$
\cdot = \cdot_0 + \cdot' + \cdot'' + \dots \quad (4.4)
$$

$$g_{\mu\nu} = \eta_{\mu\nu} + \varepsilon_{\mu\nu} = \begin{bmatrix} 1 & 0 & 0 & 0 \\ 0 & -1 & 0 & 0 \\ 0 & 0 & -1 & 0 \\ 0 & 0 & 0 & -1 \end{bmatrix} + h^{(1)}_{\mu\nu} + h^{(2)}_{\mu\nu} + ... \tag{4.5}$$

($\varepsilon_{\mu\nu}\varepsilon^{\mu\nu} = |\varepsilon|^2 << 1$) where $\eta_{\nu\mu}$ satisfies the static solution $R^{(0)}_{\mu\nu\,(\eta_{\nu\mu})} = 0, R^{(0)}_{(\eta_{\nu\mu})} = 0$, of the zero order GE

$$R^{(0)}_{\mu\nu} - \frac{1}{2} R^{(0)} = 0, \tag{4.6}$$

$\psi_0$ is the solution of the zero order KGE,

$$\partial_\mu \partial^\mu \psi_0 = -\frac{m^2 c^2}{\hbar^2} \psi_0, \tag{4.7}$$

$\psi'$ the solution of the first order KGE

$$\partial_\mu \partial^\mu \psi' + \frac{m^2 c^2}{\hbar^2} \psi' = -\partial_\mu h^{(1)\mu\nu} \partial_\nu \psi \cong -\partial_\mu h^{(1)\mu\nu} \partial_\nu \psi_0, \tag{4.8}$$

(at first order $\varepsilon_{\mu\nu} = h^{(1)}_{\mu\nu}$) and $h^{(1)}_{\mu\nu}$ is the solution of the first order GE

$$R^{(1)}_{\mu\nu\left(h^{(1)}_{\mu\nu}\right)} - \frac{1}{2} R^{(1)}_{\mu\nu\left(h^{(1)}_{\mu\nu}\right)} g_{\mu\nu} + \frac{8\pi G}{c^4} \Lambda g_{\mu\nu} = \frac{8\pi G}{c^4} |\psi|^2 Tr\left( T^{(0)}_{\mu\nu class} - \Lambda^{(0)}_Q g_{\mu\nu} + \chi T^{(0)}_{\mu\nu stress} \right) \tag{4.9}$$

where the Christoffel symbol reads [22]

$$\Gamma^{\kappa}_{\nu\mu} = \frac{1}{2} \eta^{\kappa\lambda} \left( \partial_\mu \varepsilon_{\lambda\nu} + \partial_\nu \varepsilon_{\lambda\mu} - \partial_\lambda \varepsilon_{\nu\mu} \right). \tag{4.10}$$

leading to [22]

$$R_{\mu\nu(h_{\mu\nu})} = \left( \partial_l \Gamma^l_{\mu\nu} - \partial_\nu \Gamma^l_{\mu l} + \Gamma^l_{\mu\nu} \Gamma^m_{lm} - \Gamma^m_{\mu l} \Gamma^l_{\nu m} \right). \tag{4.11}$$

Moreover, by using the zero-order relations $V_{qu0(k)} = 0$, $\kappa_{(V_{qu0(k)})} = 0$, $T^{(0)}_{(k)\mu\nu mix} = 0$, the components $T^{(0)}_{\mu\nu class}$, $\chi T^{(0)}_{\mu\nu stress}$ and $\Lambda^{(0)}_Q g_{\mu\nu}$ respectively read

$$T^{(0)}_{\mu\nu class} = \rho_{0(k)} \frac{mc^2}{\gamma} g_{\mu\kappa} u^\kappa u^\nu \delta_{hk}$$

$$= \rho_{0(k)} c^2 \frac{i\hbar}{2} g_{\mu\kappa} \frac{\frac{\partial ln[\frac{\psi_{0k}}{\psi_{0k}^*}]}{\partial q_\kappa} \frac{\partial ln[\frac{\psi_{0k}}{\psi_{0k}^*}]}{\partial q_\nu}}{\sqrt{1 - \frac{V_{qu0(k)}}{mc^2}} \frac{\partial ln[\frac{\psi_{0k}}{\psi_{0k}^*}]}{\partial t}} \delta_{hk}, \tag{4.12}$$

$$= \rho_{0(k)} c^2 g_{\mu\kappa} \frac{k^\kappa k^\nu}{\mu} \delta_{hk}$$

where

$$s_{0(k)} = s_{(k)(\cdot_{0k})}, \tag{4.13}$$

and, being $2_{\text{«¬}curv(\cdot_{0(k)})} = 0$ ,

$$a\, 2^{(0)}_{\text{«¬}stress} = -\left( \left( \begin{array}{c} \frac{1}{2} \frac{b_k |\cdot_{0k}| exp[\frac{iS_{(k)}}{\hbar}] \left( \begin{array}{c} \left( \sim_{(V_{qu_0(k)})} L_{(k)class} \Delta_{s\text{«¬}(k)} \right) \\ + 2^{(0)}_{s(k)\text{«¬}mix} \end{array} \right)}{\sum_j b_j |\cdot_j| exp[\frac{iS_{(j)}}{\hbar}]} \\ + \frac{1}{2} \frac{b^*_k |\cdot_{0k}| exp[\frac{-iS_{(k)}}{\hbar}] \left( \begin{array}{c} \left( \sim_{(V_{qu_0(k)})} L_{(k)class} \Delta_{s\text{«¬}(k)} \right) \\ - 2^{(0)}_{s(k)\text{«¬}mix} \end{array} \right)}{\sum_j b^*_j |\cdot_j| exp[\frac{-iS_{(j)}}{\hbar}]} \end{array} \right) + 2_{s\text{«¬}curv} \right) ¢_{hk} = 0 \tag{4.14}$$

$$i^{(0)}_{Q} = -\left( \left( \begin{array}{c} \frac{b_k |\cdot_{0k}| e^{[\frac{iS_{(k)}}{\hbar}]}}{2\sum_j b_j |\cdot_j| e^{[\frac{iS_{(k)}}{\hbar}]}} \left( \begin{array}{c} \sim_{(V_{qu_0(k)})} \left( \begin{array}{c} \Lambda_{class} - \frac{\Delta_{©©(k)}}{4} c^2 \frac{i\hbar}{2} g_{\sim} \\ \frac{\frac{\partial ln[\frac{\cdot_{0k}}{\cdot_{0k}^*}]}{\partial q_{\sim}} \frac{\partial ln[\frac{\cdot_{0k}}{\cdot_{0k}^*}]}{\partial q}}{\sqrt{1-\frac{V_{qu_0(k)}}{mc^2}} \frac{\partial ln[\frac{\cdot_{0k}}{\cdot_{0k}^*}]}{\partial t}} \end{array} \right) \\ + \frac{2^{(0)}_{(k)©©\,mix}}{4} \end{array} \right) \\ + \frac{b^*_k |\cdot_{0k}| e^{[-\frac{iS_{(k)}}{\hbar}]}}{2\sum_j b^*_j |\cdot_j| e^{[-\frac{iS_{(k)}}{\hbar}]}} \left( \begin{array}{c} \sim_{(V_{qu_0(k)})} \left( \begin{array}{c} \Lambda_{class} - \frac{\Delta_{©©(k)}}{4} c^2 \frac{i\hbar}{2} g_{\sim} \\ \frac{\frac{\partial ln[\frac{\cdot_{0k}}{\cdot_{0k}^*}]}{\partial q_{\sim}} \frac{\partial ln[\frac{\cdot_{0k}}{\cdot_{0k}^*}]}{\partial q}}{\sqrt{1-\frac{V_{qu_0(k)}}{mc^2}} \frac{\partial ln[\frac{\cdot_{0k}}{\cdot_{0k}^*}]}{\partial t}} \end{array} \right) \\ - \frac{2^{(0)}_{(k)©©\,mix}}{4} \end{array} \right) \end{array} \right) + \frac{2_{aa\,curv}}{4} \right) ¢_{hk} = 0 \tag{4.15}$$

leading to the first-order GE

$$R_{\text{«¬}\left(h^{(1)}_{\text{«¬}}\right)} - \frac{1}{2} R_{\left(h^{(1)}_{\text{«¬}}\right)} g_{\text{«¬}} + \frac{8®G}{c^4} \Lambda g_{\text{«¬}} = \frac{8®G}{c^4} Tr\left( 2^{(0)}_{\text{«¬}class} \right). \tag{4.16}$$

By making the macroscopic limit of (4.16) (with $\Lambda = 0$) we obtain

$$lim_{macro}\left( R_{«¬\left(h^{(1)}_{«¬}\right)} - \frac{1}{2} R_{\left(h^{(1)}_{«¬}\right)} g_{«¬} \right) = lim_{\hbar \to 0} \frac{8®G}{c^4} |\cdot_{\tilde{k}}|^2 \frac{mc^2}{¥} g_{«\sim} u^{\sim} u^{¬}$$

$$R^{macro}_{«¬\left(h^{(1)}_{«¬}\right)} - \frac{1}{2} R^{macro}_{\left(h^{(1)}_{«¬}\right)} g_{«¬} = \frac{8®G}{c^4} |\cdot|^2 \frac{mc^2}{¥} g_{«\sim} u^{\sim}_{class} u^{¬}_{class} \quad . \qquad (4.17)$$

from which we can readily see that the weak gravity limit, on macroscopic scale at the first-order, leads to the Newtonian potential of gravity. The first contribution to the cosmological constant comes from the second order of approximation

$$R^{(2)}_{«¬\left(£_{«¬}\right)} - \frac{1}{2} R^{(2)}_{\left(£_{«¬}\right)} g_{«¬} = \frac{8®G}{c^4} |\cdot|^2 \, Tr\left( {}^{2}{}^{(1)}_{«¬ class} - ¡^{(1)}_{Q} g_{«¬} + a^{2}{}^{(1)}_{«¬ stress} \right) \qquad (4.18)$$

where $£_{«¬} = h^{(1)}_{«¬} + h^{(2)}_{«¬}$ and the EITD is calculated by using $\cdot\,'$ obtained by (4.4), where

$$¡^{(1)}_{Q} = -\left( \left( \frac{b_k |\cdot_k| e^{[\frac{iS(k)}{\hbar}]}}{2\sum_j b_j |\cdot_j| e^{[\frac{iS(k)}{\hbar}]}} \left( \sim_{(V_{qu(k)})} \left( \frac{ \Lambda - \frac{\Delta_{aa(k)}}{4} c^2 \frac{i\hbar}{2} g_{\sim} \frac{\partial ln[\frac{\cdot_k}{\cdot_k *}]}{\partial q_{\sim}} \frac{\partial ln[\frac{\cdot_k}{\cdot_k *}]}{\partial q} }{ \sqrt{1 - \frac{V_{qu(k)}}{mc^2}} \frac{\partial ln[\frac{\cdot_k}{\cdot_k *}]}{\partial t} } \right) + \frac{{}^{2}{}_{(k)aa\ mix}}{4} \right) + \frac{b*_k |\cdot_k| e^{[-\frac{iS(k)}{\hbar}]}}{2\sum_j b*_j |\cdot_j| e^{[-\frac{iS(k)}{\hbar}]}} \left( \sim_{(V_{qu(k)})} \left( \frac{ \Lambda - \frac{\Delta_{aa(k)}}{4} c^2 \frac{i\hbar}{2} g_{\sim} \frac{\partial ln[\frac{\cdot_k}{\cdot_k *}]}{\partial q_{\sim}} \frac{\partial ln[\frac{\cdot_k}{\cdot_k *}]}{\partial q} }{ \sqrt{1 - \frac{V_{qu(k)}}{mc^2}} \frac{\partial ln[\frac{\cdot_k}{\cdot_k *}]}{\partial t} } \right) - \frac{{}^{2}{}_{(k)aa\ mix}}{4} \right) \right) + \frac{{}^{2}{}_{aa\ curv}}{4} \right) ¢_{hk} \qquad (4.19)$$

where $\cdot_k = \cdot_{0k} + \cdot'_k$ .

Moreover, being $lim_{decoherence}\, {}^{2}{}_{«¬ curv} = 0$, the decoherent (macroscopic) limit (with $\Lambda = 0$) of the GE reads

$$R^{(2)dec}_{«¬\left(£_{«¬}\right)} - \frac{1}{2} R^{(2)dec}_{\left(£_{«¬}\right)} g_{«¬} = \frac{8®G}{c^4} |\cdot_{\tilde{k}}|^2 \left( \left(1 - \sim^{(1)}_{(\tilde{k})}\right) \frac{mc^2}{¥} g_{«\sim} u^{\sim} u^{¬} + \Lambda^{(1)dec}_{(\tilde{k})Q} g_{«¬} + \Delta^{2}{}^{(1)dec}_{«¬ stress} \right) \qquad (4.20)$$

where

$$\frac{mc^2}{\text{¥}} u^{\sim} u^{\neg} = c^2 \frac{i\hbar}{2} \frac{\frac{\partial ln[\frac{\cdot_k}{\cdot_k{}^*}]}{\partial q_{\sim}} \frac{\partial ln[\frac{\cdot_k}{\cdot_k{}^*}]}{\partial q_{\neg}}}{\sqrt{1-\frac{V_{qu(k)}}{mc^2}} \frac{\partial ln[\frac{\cdot_k}{\cdot_k{}^*}]}{\partial t}} = \frac{mc^2}{\text{¥}}\left(u^{\sim}_{class}u^{\neg}_{class} + \Delta u^{\sim}\Delta u^{\neg}\right), \qquad (4.21)$$

where

$$\frac{mc^2}{\text{¥}} u^{\sim}_{class} u^{\neg}_{class} = c^2 \frac{i\hbar}{2} \frac{\frac{\partial ln[\frac{\cdot_{0k}}{\cdot_{0k}{}^*}]}{\partial q_{\sim}} \frac{\partial ln[\frac{\cdot_{0k}}{\cdot_{0k}{}^*}]}{\partial q_{\neg}}}{\sqrt{1-\frac{V_{qu_0(k)}}{mc^2}} \frac{\partial ln[\frac{\cdot_{0k}}{\cdot_{0k}{}^*}]}{\partial t}} = c^2 \frac{k^{\sim} k^{\neg}}{\mu_k} \qquad (4.22)$$

where, at first order,

$$\begin{aligned} \frac{mc^2}{\text{¥}} \Delta u^{\sim} \Delta u^{\neg} \cong\ & c^2 \frac{k^{\sim} k^{\neg}}{\mu_k{}^2} \left( -\partial_t \left( \frac{\cdot'_k}{\cdot_{0k}} - \frac{\cdot'_k{}^*}{\cdot_{0k}{}^*} \right) + \frac{\mu_k V^{(1)}_{qu(\cdot'_k)}}{2mc^2} \right) \\ & + c^2 \frac{i\hbar}{2\mu_k} \left( k^{\sim} \partial^{\neg} \left( \frac{\cdot'_k}{\cdot_{0k}} - \frac{\cdot'_k{}^*}{\cdot_{0k}{}^*} \right) + \partial^{\sim} \left( \frac{\cdot'_k}{\cdot_{0k}} - \frac{\cdot'_k{}^*}{\cdot_{0k}{}^*} \right) k^{\neg} \right) \end{aligned}, \qquad (4.23)$$

and where

$$\begin{aligned} \Lambda^{(1)dec}_{(k)\ \ Q} &= -\sim^{(1)}_{(V_{qu_1(k)})} \Delta^{(1)}_{©©(k)} g_{\sim}\ \ c^2 \frac{i\hbar}{8} \frac{\frac{\partial ln[\frac{\cdot_k}{\cdot_k{}^*}]}{\partial q_{\sim}} \frac{\partial ln[\frac{\cdot_k}{\cdot_k{}^*}]}{\partial q}}{\sqrt{1-\frac{V_{qu\ (k)}}{mc^2}} \frac{\partial ln[\frac{\cdot_k}{\cdot_k{}^*}]}{\partial t}} \\ &= -\sim^{(1)}_{(V_{qu_1(k)})} \frac{\Delta^{(1)}_{©©(k)}}{4} g_{\sim}\ \ \frac{mc^2}{\text{¥}} u^{\sim} u \\ &= -\sim^{(1)}_{(V_{qu_1(k)})} \frac{\Delta^{(1)}_{©©(k)}}{4} g_{\sim}\ \ \frac{mc^2}{\text{¥}} \left( u^{\sim}_{class} u_{class} + \Delta u^{\sim} \Delta u \ \ \right) \\ &\cong -\sim^{(1)}_{(V_{qu_1(k)})} \frac{\Delta^{(1)}_{©©(k)}}{4} g_{\sim}\ \ \frac{mc^2}{\text{¥}} u^{\sim}_{class} u_{class} \\ &\cong -\sim^{(1)}_{(V_{qu_1(k)})} \Delta^{(1)}_{©©(k)} g_{\sim}\ \ c^2 \frac{k^{\sim} k}{\mu_k} \end{aligned} \qquad (4.24)$$

and

$$\Delta^{2\,(1)dec}_{«¬stress} = -\sim^{(1)}_{(V_{qu1(k)})} \Delta^{(1)}_{s«¬(k)} g_{\sim} \; c^2 \frac{i\hbar}{2} \frac{\frac{\partial ln[\frac{\cdot_k}{\cdot_k^*}]}{\partial q_{\sim}} \frac{\partial ln[\frac{\cdot_k}{\cdot_k^*}]}{\partial q}}{\sqrt{1-\frac{V_{qu\ (k)}}{mc^2}} \frac{\partial ln[\frac{\cdot_k}{\cdot_k^*}]}{\partial t}}$$

$$= -\sim^{(1)}_{(V_{qu1(k)})} \Delta^{(1)}_{s«¬(k)} g_{\sim} \; \frac{mc^2}{¥} u^{\sim} u$$

$$= -\sim^{(1)}_{(V_{qu1(k)})} \Delta^{(1)}_{s«¬(k)} g_{\sim} \; \frac{mc^2}{¥} \left( u^{\sim}_{class} u_{class} + \Delta u^{\sim} \Delta u \right) \quad (4.25)$$

$$\cong -\sim^{(1)}_{(V_{qu1(k)})} \Delta^{(1)}_{s«¬(k)} g_{\sim} \; \frac{mc^2}{¥} u^{\sim}_{class} u_{class}$$

$$\cong -\sim^{(1)}_{(V_{qu1(k)})} \Delta^{(1)}_{s«¬(k)} g_{\sim} \; c^2 \frac{k^{\sim} k}{\mu_k}$$

If the macroscopic GE (4.17) and the Einstein equation of the general relativity coincide themselves at first order and the Newtonian gravity is purely classic, the second order GE (4.20) contains contributions (among those the cosmological isotropic pressure $\Lambda^{(1)}_{(k)Q}$ ) that go to zero if $\hbar$ is set to zero so that $\Lambda^{(1)}_{(k)Q}$ actually is the macroscopic *quantum-mechanical* contribution to the Newtonian gravity.

It is worth mentioning that the macroscopic GE shows the additional contribution $\Delta^{2\,(1)dec}_{«¬stress}$ to the cosmological isotropic pressure. The dependence of such term by both $\sim^{(1)}_{(V_{qu1(k)})}$ and $\Delta^{(1)}_{s«¬(k)}$, that become relevant in very high-curvature space-time, suggests that this term gives detectable effects near the big black holes at the center of the galaxies.

It is noteworthy that the EIT stress component $a^{2}_{«¬stress}$, that leads to a non-zero slip function [23-24], is specific of the (microscopic) quantum-coherent curved space-time but it decays to $\Delta^{2\,(1)dec}_{«¬stress}$ for the decoherence at the macroscopic scale.

## 5. The CPTD expectation-value of the quantum KGE field

Generally speaking, when the KGE field is quantized, the EITD on the right side of the GE becomes a quantum operator and thence also the Ricci's tensor (as well as the metric tensor) of the GE, on the left side, become quantum operators. As it can be easily shown for the pure gravity [25], the commutating rules for the KGE field quantization fixes the commuting relations for the metric tensor.

At zero order, the GE equation leads to a Minkowskian KGE field and, hence, when the KGE field is quantized, the standard QFT outputs are obtained.

If at zero order the GE is decoupled by the field of the massive KGE, at higher order it is not.

The quantization of the GE-KGE system of equations is not the goal of this work, nevertheless, it is interesting to evaluate the CPTD expectation value of the vacuum in order to evaluate if it can lead to the lowering of the theoretical value of the CC on cosmological scale and can help to solve the problem of the disagreement of the QFT with the experimental observations.

In order to evaluate the macroscopic cosmological constant of the quantum KGE field (i.e., at the zero order Minkowskian limit of the GE-KGE system of equations for the ordinary QFT) we need to calculate the expectation value $< 0_k \,|lim_{decoherence}\; i_Q\, g_{\mu\nu} \,|0_k> = < 0_k \,|\Lambda^{(1)dec}_{(k)Q}\, g_{\mu\nu} \,|0_k>$. To this end, we need to express the quantum potential as a function of the annihilation and creation operators $a_{(k)}$ and $a^{\dagger}_{(k)}$ of the Fourier decomposition of the free KGE quantum field

$$\phi = \iint \frac{d^3k}{(2\pi)^3} \frac{1}{2\omega_k} \left( a_{(k)} exp[ik_\mu q^\mu] + a^{\dagger}_{(k)} exp[-ik_\mu q^\mu] \right) , \qquad (5.1)$$

that, by using the discrete form of field Fourier decomposition (i.e., $\iint \frac{d^3k}{(2\pi)^3} \to \frac{1}{\sqrt{V}} \sum_{k=0}$ ) reads

$$\phi = \frac{1}{\sqrt{V}} \sum_{k=0}^{\infty} \frac{1}{2\omega_k} \left( a_{(k)} exp[\frac{ip_\mu q^\mu}{\hbar}] + a^{\dagger}_{(k)} exp[-\frac{ip_\mu q^\mu}{\hbar}] \right) \qquad (5.2)$$

where $\frac{p_\mu}{\hbar} = k_\mu$, where the identities

$$a(k) \equiv 2\omega_k b(k) |\phi_k| \qquad (5.3)$$

and

$$a^{\dagger}(k) \equiv 2\omega_k b(-k) |\phi_{-k}| \qquad (5.4)$$

can be established with notation in (2.1.1) (where for $k < 0 \Rightarrow$ both $a(k) \to a^{\dagger}(k)$ and $a^{\dagger}(k) \to a(k)$) and leads to

$$\begin{aligned} V_{qu0(k>0)} &= -\frac{\hbar^2}{m} \frac{1}{|\phi_k| \sqrt{-g}} \partial_\mu \sqrt{-g} g^{\mu\nu} \partial_\nu \,|a(k) exp[ik_\mu q^\mu]| \\ &= -\frac{\hbar^2}{m} \frac{1}{|\phi_k|} \partial_\mu \partial^\mu \,|a(k) exp[ik_\mu q^\mu]| \\ &= -\frac{\hbar^2}{m} \frac{1}{|\phi_k|} \partial_\mu \partial^\mu \sqrt{a(k) a^{\dagger}(k)} = V_{qu0(k<0)} \end{aligned} \qquad (5.5)$$

where, now, $a(k), a^{\dagger}(k)$ are quantum operators obeying to the commutation relations

$$[a(k), a^{\dagger}(k')] = \delta_{kk'} . \qquad (5.6)$$

However, even if the squared root of operators in (5.5) in the quantum potential can be defined by making use of the Taylor expansion series, the higher order terms of the such expansions own the ordering problem of the quantum operators. In order to remark this freedom in the definition of the quantum potential operator, we name it as $\hat{V}_{qu}^{Q-ord}$ and reads

$$\hat{V}_{qu}^{Q-ord} = -\frac{\hbar^2}{m} \frac{1}{\left(|\phi_k|\right)^{Q-ord} \sqrt{-g}} \partial_\mu \sqrt{-g} g^{\mu\nu} \partial_\nu \left(|\phi_k|\right)^{Q-ord} \qquad (5.7)$$

leading to the Minkowskian limit

$$\hat{V}_{qu0}^{Q-ord} = -\frac{\hbar^2}{m} \frac{1}{\left(\sqrt{a(k) a^{\dagger}(k)}\right)^{Q-ord} \sqrt{-g}} \partial_\mu \sqrt{-g} g^{\mu\nu} \partial_\nu \left(\sqrt{a(k) a^{\dagger}(k)}\right)^{Q-ord} \qquad (5.8)$$

.

Moreover, by using (2.0.9) and the identities $p_\mu = -\partial_\mu S = (p_0, -p_i)$ and $p^2 = p_i p_i$, it follows that

$$ \mu_k{}^2 = \frac{c^2 p^2}{\hbar^2} + \frac{m^2 c^4}{\hbar^2}\left( 1 - \frac{\hat{V}_{qu}^{Q-ord}}{mc^2} \right). \qquad (5.9) $$

Moreover, by using (3.2.4), and being $lim_{decoherence}{}^{2}«¬_{mix} = 0$ and $lim_{decoherence}{}^{2}«¬_{curv} = 0$, the macroscopic decoherent CPTD

$$ lim_{decoherence}\ \tilde{\text{ĩ}}_{Q} = \tilde{\Lambda}_{(k)Q} = -\sim_{(\hat{V}_{qu0}^{Q-ord})}\left( -\frac{\Delta_{©©(k)}}{4} c^2 g_{\sim}\ \frac{k_k}{\mu_k \sqrt{1 - \frac{\hat{V}_{qu0}^{Q-ord}}{mc^2}}} \right) \qquad (5.10) $$

whose expectation value reads

$$ < 0_k \,|\, \tilde{\Lambda}_{(k)Q} \,|\, 0_k > = \frac{1}{(2®)^2 \hbar c} \int^{\mu_{k_{max}}} \tilde{\Lambda}_{(k)Q}^{(0)dec} \sqrt{\frac{\hbar^2 \mu_k{}^2}{c^2} - m^2 c^2}\, d\mu_k \qquad (5.11) $$

where $|0_k>$ represents the k-indexed harmonic oscillators of the field in the fundamental state [19].

Moreover, being in the Minkowskian limit $\partial^{«}\left(\sqrt{a(k)a^{\dagger}(k)}\right)^{Q-ord} = 0$, it follows that

$$ \hat{V}_{qu0}^{Q-ord} = 0 \qquad (5.12) $$

$$ \sim_{(\hat{V}_{qu0}^{Q-ord})} = 0, \qquad (5.13) $$

to

$$ \Delta_{©©(k)} = 0 \qquad (5.14) $$

to

$$ \tilde{\Lambda}_{(k)Q} = 0, \qquad (5.15) $$

to

$$ < 0_p \,|\, \tilde{\Lambda}_{(k)Q} \,|\, 0_p > = 0. \qquad (5.16) $$

and, finally, from (5.12), to

$$ \mu_k{}^2 = \frac{c^2 p^2}{\hbar^2} + \frac{m^2 c^4}{\hbar^2} = c^2 k^2 + \frac{m^2 c^4}{\hbar^2}. \qquad (5.17) $$

From the last identity, the standard QFT outputs are warranted at the zero order of approximation.

## 6. Discussion

The hydrodynamic representation of the KGE field makes it equivalent to a mass distribution $|\cdot|^2$ submitted to the non-local quantum potential that leads to the GE (3.2.1).

The basic assumption of the presented theory is that such hydrodynamic representation of the KGE field (restricted to irrotational states) owns a physical reality since, as shown by the Aharonov-Bohm effect, the quantum potential can be experimentally detected. Therefore, for the basic principle of the general relativity, the (kinetic) energy of the quantum potential contributes to the curvature of the space-time. On the basis of this postulate, the quantum-mechanical non-local effects come into the gravity leading to the theoretical appearance of the CPTD $\text{ì}_{Q}$ in the GE.

In the classical treatment, the Einstein equation for massive particles is not coupled to any field, but just to the energy impulse tensor of classical bodies and does not contain any information about how fields couple with it.

The GE (3.2.1) is analytically coupled to the KGE field. The GE-KGE system can be further quantized leading to a quantum gravity theory derived by an analytically field-defined Einstein-Hilbert action.

### 6.1. Analogy with the Brans-Dicke gravity

The output the work highlights an interesting analogy with the Brans-Dicke [13] gravity that solves the problem of the cosmological constant [26] as well as those of the inflaction [27] and dark energy [28]. If we look in detail to the EITD $|\cdot|^{2}\,{}^{2}{}_{«¬class}$ (3.2.2), the macroscopic limit

$$lim_{macro}|\cdot|^{2}\,{}^{2}{}_{«¬class} = T_{(\tilde{k})«¬class}, \qquad (6.1.1)$$

leads to

$$\begin{aligned}\frac{8®G}{c^4}T_{«¬class} &= \frac{8®G}{c^4}|\cdot|^2 c^2\left(\frac{\partial S}{\partial t}\right)^{-1}\left(g^{«}\frac{\partial S}{\partial q}\frac{\partial S}{\partial q^{¬}}\right)\\ &= -\frac{8®G}{c^4}|\cdot|^2\frac{\hbar^2c^2}{4E}\left(g^{«}\left(\frac{\partial\ \cdot}{\cdot}-\frac{\partial\ \cdot *}{\cdot *}\right)\left(\frac{\partial_{¬}\cdot}{\cdot}-\frac{\partial_{¬}\cdot *}{\cdot *}\right)\right)\\ &= -\frac{8®}{c^4}\frac{G}{\cdot^{2}}|\cdot|^2\frac{\hbar^2c^2}{4E}\left(g^{«}\partial\ \cdot\partial_{¬}\cdot+........+\left(\frac{\cdot}{\cdot *}\right)^2 g^{«}\partial\ \cdot *\partial_{¬}\cdot *\right)\end{aligned} \qquad (6.1.2)$$

showing to be composed by terms that are of the form of those contained into the Brans-Dicke gravity [13] in absence of external potentials.

If is worth noting that the hydrodynamic gravity, gives theoretical support to the Brans-Dicke effective gravitational constant $G_{eff} = \frac{G}{\cdot^{2}}$ .

### 6.2. The GE and the quantum gravity

Even if the quantization of the KGE field in the curved space defined by the GE (3.2.1) is not treated in this work, the inspection of some features of quantum gravity to the light of the GE (3.2.1) deserves a mention.

Since the action of the GE (3.2.1) is basically given by the standard Einstein-Hilbert action plus terms stemming by the energy of the non-local quantum potential of massive KGE, the outputs of the quantum "pure gravity" practically remains almost valid.

As shown in [10], one interesting aspect of the quantum pure gravity is that the vacuum does not make a transition to the *collapsed branched polymer phase*, if an even small cosmological constant (i.e., $\Lambda_{class} \neq 0$) is present. Respect to this fact, the presence of the term $\Lambda_Q$ in the GE (??), in principle, allows to work with the assumption $\Lambda_{class} = \Lambda = 0$ with the GE that reads.

$$R_{«¬} - \frac{1}{2}Rg_{«¬} + \frac{8®G}{c^4}Tr\left(\tilde{\mathrm{i}}_{Q}\right)g_{«¬} = \frac{8®G}{c^4}Tr\left({}^{2}{}_{«¬class} + \mathrm{a}^{2}{}_{«¬stress}\right) \qquad (6.2.1)$$

Moreover, since the matter itself makes the space-time curved and, hence, $\sim_{(V_{qu})} \neq 0$, $\mathrm{a}^{2}{}_{«¬stress} \neq 0$ and

$$
\tilde{\text{ì}}^{\,Q-ord}_{Q(k>0\square\,,k<0\,\square)} = -\frac{1}{2}\left(\left(\begin{array}{l}
\dfrac{\dfrac{e^{[\frac{iS_{(k)}}{\hbar}]}}{\mu_k}\ \begin{matrix}\square\ a_k\\ \square\ a_k^{\dagger}\end{matrix}}{\sum_{k'=0}^{\infty}\dfrac{a_{k'}}{\mu_{k'}}e^{[\frac{iS_{(k')}}{\hbar}]}+\dfrac{a^{\dagger}{}_{k'}}{\mu_{k'}}e^{[\frac{iS_{(-k')}}{\hbar}]}}\left(\sim\left(\dfrac{-\dfrac{\Delta_{\text{©©}(k)}}{4}c^2 g_{\sim}\dfrac{\partial S_{(k)}}{\partial q_{\sim}}\dfrac{\partial S_{(k)}}{\partial q}}{\sqrt{1-\dfrac{V_{qu(k)}}{mc^2}\dfrac{\partial S_{(k)}}{\partial t}}}\right)+\dfrac{{}^{2}{}_{(k)\text{©©}\,mix}}{4}\right)\\
+\dfrac{\dfrac{e^{[-\frac{iS_{(k)}}{\hbar}]}}{\mu_k}\ \begin{matrix}\square\ a_k\\ \square\ a_k^{\dagger}\end{matrix}}{\sum_{k'=0}^{\infty}\dfrac{a^{\dagger}{}_{k'}}{\mu_{k'}}e^{[-\frac{iS_{(k')}}{\hbar}]}+\dfrac{a_{k'}}{\mu_{k'}}e^{[-\frac{iS_{(-k')}}{\hbar}]}}\left(\sim\left(\dfrac{-\dfrac{\Delta_{\text{©©}(k)}}{3}c^2 g_{\sim}\dfrac{\partial S_{(k)}}{\partial q_{\sim}}\dfrac{\partial S_{(k)}}{\partial q}}{\sqrt{1-\dfrac{V_{qu(k)}}{mc^2}\dfrac{\partial S_{(k)}}{\partial t}}}\right)-\dfrac{{}^{2}{}_{(k)\text{©©}\,mix}}{4}\right)
\end{array}\right)+\frac{{}^{2}{}_{\text{aa}\,curv}}{4}\right)^{Q-ord}\ \text{¢}_{hk}\ ,
$$

(6.2.2)

it follows that the matter itself stabilizes the vacuum in the physical strong gravity phase [10] as we perceive it.

On the other hand, under this hypothesis, a perfect Minkowskian vacuum (i.e., without matter) will make transition to the unphysical *collapsed branched polymer phase* with no sensible continuum limit [10-11], leading to no-space and no-time as we experience.

### 6.3. Check of the hydrodynamic GE

If in the general relativity the energy-impulse tensor density for classical bodies [30] is defined only with a point-dependence by the mass density, for the electromagnetic (EM) field, the EITD is defined as a function of the EM field itself [22].

On this basis, since the photon is a boson (obeying the a KGE), we can have a direct check of the theory by comparing the known EITD EM expression

$$
T_{\text{¬«}\,em} = \frac{1}{4\text{®}}\left(-F_{\text{«a}}F_{\text{¬}}{}^{\text{a}} + \frac{1}{4}F_{\text{a¥}}F^{\text{a¥}}g_{\text{«¬}}\right) \qquad (6.3.1)
$$

with the EITD (3.6) for a boson field given by the quantum hydrodynamic gravity.
In fact, given the plane wave of the vector potential, in the Minkowskian case, for the photon (e.g. linearly polarized)

$$
A_k = A_0\ exp[-ik_{\text{«}}q^{\text{«}}] = A_{0\text{¥}}\ exp[i\frac{S_{(k)}}{\hbar}] \qquad (6.3.2)
$$

being $|\cdot_k|^2$ the number of particles (i.e., photons) per volume, the KGE field $\cdot_k$ reads

$$\cdot_k \propto \frac{A_k}{c}\sqrt{\frac{\mu}{\hbar}} \quad \left[ l^{-3/2} \right], \tag{6.3.3}$$

and, being $V_{qu(k)} = 0$ , $\sim_{(V_{qu(k)})} = 0$, and $L = -g_{«\sim} k^{\sim} k^{«} = 0$, we obtain

$$^{2}{}_{«¬class} = T_{(\tilde{k})«¬class} , \tag{6.3.4}$$

$$\text{i}_{«¬Q} = -\frac{|\cdot_k|^2}{2}\left( \frac{^{2}{}_{(k)©©\,mix}}{4} - \frac{^{2}{}_{(k)©©\,mix}}{4} \right) g_{«¬} ¢_{hk} = 0 , \tag{6.3.5}$$

$$\text{a}^{2}{}_{«¬stress} = -\frac{|\cdot_k|^2}{2}\left( ^{2}{}_{s(k)«¬mix} - ^{2}{}_{s(k)«¬mix} \right) ¢_{hk} = 0 \tag{6.3.6}$$

from which the EITD for the photon reads

$$T_{«¬} = T_{(k)«¬class} = |\cdot_k|^2\, \text{T}_{(k)«¬class}$$
$$= |\cdot_k|^2 c^2 \left( \frac{\partial S_{(k)}}{\partial t} \right)^{-1} \left( \frac{\partial S_{(k)}}{\partial q^{«}} \frac{\partial S_{(k)}}{\partial q^{¬}} - g_{\sim} \frac{\partial S_{(k)}}{\partial q} \frac{\partial S_{(k)}}{\partial q^{\sim}} g_{«¬} \right) \propto |A|^2 k_{«} k_{¬} , \tag{6.3.7}$$

that, compared to the output of (6.3.1) for the photon [31]

$$T_{¬«\ em} = \frac{|E|^2}{4®}\frac{c^2}{\mu^2} k_{«} k_{¬} = \frac{|A|^2}{4®} k_{«} k_{¬} \quad , \tag{6.3.8}$$

leads to

$$T_{«¬} \propto 4® T_{¬«\ em} \tag{6.3.9}$$

### 6.4. Experimental tests

The result (5.15-6) basically shows that in the Minkowskian space-time (i.e., the vacuum very far from particles) the CPTD expectation value is null regardless its zero-point energy density.

The non-zero contribution to the CC will appear at second order in the GE deriving by the first order $\cdot'$ of the quantized KGE field. In qualitative way, the CPTD macroscopic expectation value is vanishing in the region of space-time with Newtonian gravity and it increases at higher gravity as quantum-mechanical corrections. This fact well agrees [19] with the very small value of the observed CC and leads to a scenario where the major contribution to the CC comes from black holes (this is due to the high mass density of a black hole where the matter is so squeezed that the quantum potential energy becomes comparable with the mass energy itself so that $V_{qu} \approx mc^2$ [29] and $\sim_{(V_{qu(k)})} \approx max\{\sim_{(V_{qu(k)})}\} \approx 1$, the *decoherent quantum-mechanical corrections* to the Newtonian gravity $\sim^{(1)}_{(\tilde{k})} \frac{mc^2}{¥} g_{«\sim} u^{\sim}_{class} u^{¬}_{class}$ , $\Lambda^{(1)dec}_{(\tilde{k})Q} g_{«¬}$ and $\Delta^{2\,(1)dec}_{«¬stress}$ in (4.20) must be primarily detectable in the motion of stars around the big black holes at the center of the galaxies., in the rotation of twin neutron stars and in the inter-galactic interaction ..

Finally, it is worth mentioning that the interferometric detection of the gravitational waves represents an experimental technique whose angular and frequency-dependent response functions can discriminate among the existing theories of gravity [32].

## 7. Conclusion

The quantum hydrodynamic representation of the Klein-Gordon equation, describing the evolution of the mass density $|\cdot|^2$ owing the hydrodynamic moment $\partial_{«} S = -p_{«}$ and subject to the quantum potential, has been used to derive the correspondent gravity equation, defining the geometry of the space-time, by using the minimum action principle. The gravity equation associated to the KGE field takes into account the gravitational effects of the energy of the non-local quantum potential.

The hydrodynamic approach owns three main properties:

1. The energy-impulse tensor of the GE is written as a function of the KGE field;
2. In the classical limit, the GE leads to the Einstein one;
3. If we apply the EITD of the GE to the photon field (that is a boson described by the KGE) we obtain the EITD of the EM theory;

The self-generation of the CPTD $Tr\left( \mathrm{i}_{Q} \right)$ leads to the attractive hypothesis that the matter itself generates the physical stable vacuum phase in which is embedded.

The paper shows that the macroscopic CPTD $\tilde{\Lambda}_{(k)Q}$ it is not null if, and only if, the space-time is curvilinear (due to the presence of localized mass) and it tends to zero in the very far flat vacuum regardless the zero-point energy of the vacuum.

The depletion of the CPTD in the vacuum, far from material bodies, lowers its mean value on cosmological scale so that it can possibly agree with the astronomical observations on the motion of the galaxies.

The GE of the classical KGE field shows that the CPTD $\tilde{\Lambda}_{(k)Q}$ and other out-diagonal components of the EITD can be considered as "decoherent quantum-mechanical" gravitational effects generated in highly curved space-time near dense matter such as black holes and neutron stars.

The hydrodynamic gravity model defines a coupling between the boson field of the free KGE and the GE in a form that, at some extend, mimics the Brans-Dicke gravity leading to an effective gravitational constant inversely proportional to the field squared.

## Appendix A

*Equivalence between the field and the hydrodynamic solutions of a quantum equation*

In addition to the field solution

$$\cdot = |\cdot| \, exp \frac{i}{\hbar} S \qquad \text{(A.0.1)}$$

of a quantum equation, it is possible to express it in the hydrodynamic form.

This approach was firstly proposed by Madelung [13] and confirmed by the Aharonov-Bohm effect [12].

In order to obtain the hydrodynamic form of a quantum equation, for the complex field (A.0.1), it is split into two equations regarding its real and imaginary part, leading to two differential equations as a function of $|\cdot|$ and $-\frac{\partial S}{\partial q^{«}}$.

With the help of the so called quantum pseudo potential [11-14], the hydrodynamic approach describes the evolution of the particles density $n = |\cdot|^2$ moving with the impulse field $p_{«}$, given by the identity

$$p_{«} = (\frac{E}{c}, -p_i) = -\frac{\partial S}{\partial q^{«}}, \qquad \text{(A.0.2)}$$

and subject to the quantum potential [11-14]. Actually, the hydrodynamic system of equations broaden the solutions of the starting quantum equation since not all momenta $p_{«}$, solutions of the hydrodynamic description, are solutions of the field equation [11]. This because not all momenta deriving from hydrodynamic equations can be obtained as the gradient of a function representing the action $S$ of the exponential argument of the field.
The restriction of the hydrodynamic momenta to those ones of the quantum problem derives by imposing the existence of the action S. The integrability of the impulse requires [11]

$$\oint p_{«} dl^{«} = -\oint \frac{\partial S}{\partial q^{«}} dl^{«} = 0 \qquad \text{(A.0.3)}$$

that for systems, that do not depend explicitly by time, is satisfied by the irrotational condition

$$\nabla \times p = 0 \qquad \text{(A.0.4)}$$

Moreover, since the action is contained in the exponential argument of the wave-function, all the multiples of $2®\hbar$, with

$$S_{n\,(q,t)} = S_{0(q,t)} + 2n®\hbar \; = \; S_{0(q_0,t)} + \int_{q_0}^{q} dl \bullet \nabla S + 2n®\hbar \qquad n = 0,\ 1,\ 2,\ 3,\ ... \qquad \text{(A.0.5)}$$

are possible, so that the action results quantized.
With the "irrotational" condition the hydrodynamic approach becomes equivalent to the field one.
Even this approach is fully quantum [11-14] and provides outputs that completely overlap those ones of the standard quantum treatment [28], its use is curtailed to the semiclassical approximation and to one-particle problem.

*Implementation of quantization in the hydrodynamic equations*

The quantization condition, that reduces the hydrodynamic solutions to those of the field of the quantum equation, as formulated by (A.0.4) is external to the hydrodynamic quantum equations (HQE).
Actually, under inspection, it is implicitly included into the HQE through the quantum potential.
In fact, in the hydrodynamic description, the eigenstates are identified by their property of stationarity that is given by the "equilibrium" condition that (for unidimensional systems) reads

$$\dot{p} = 0 \cdot \qquad \text{(A.0.6)}$$

This equilibrium, in bounded systems, happens when the force generated by the quantum potential exactly counterbalances that one of the Hamiltonian potential (see, further on, relation (A.1.10)). For 3-D systems, the equilibrium condition also refers to the rotational "equilibrium "(i.e., angular momentum conservation) in a way that the overall "torque" due both to the Hamiltonian and to the quantum potential is null.
The initial condition

$$\dot{q} = \dot{q}_0 \qquad \text{(A.0.7)}$$

united to the equilibrium condition (A.0.6) leads to

$$\dot{q}_0 = constant\,. \qquad \text{(A.0.8)}$$

that in presence of an external potential, not function of time, can be set to $\dot{q}_0 = 0$ by the choice of an opportune reference system.
It is straightforward to check by inserting (A.0.6) (i.e., $p = constant$ ) into (A.0.4) that the eigenstates are irrotational.

Moreover, since the quantum potential is not fixed but changes with the state of the system, more than one stationary state (each one with its own $V_{qu_n}$) may exist. In this case, we have multiple quantized action values $S_{n\,(q,t)}$ for each $V_{qu_n}$. (see section A.2-4).

Thence, given that the stationary states, generated by the quantum potential, are the field eigenstates, the quantization is implicitly-contained in the quantum hydrodynamic equations.

*A.1.Simple examples*

In this section, in order to elucidate how the quantum potential defines the eigenstates, some simple cases are reported. The skilled reader may skip these paragraphs.

Here we consider the low velocity limit of the KGE submitted to the potential $V_{(q)}$ at first order in $\frac{|\dot{q}|}{c}$ (i.e., the Schrödinger equation) for which it holds

$$
\begin{aligned}
& lim_{\frac{|\dot{q}|}{c}\to 0} L_k - V_{(q)} = -lim_{\frac{|\dot{q}|}{c}\to 0} \frac{mc^2}{¥}\sqrt{1-\frac{V_{qu(k)}}{mc^2}} - V_{(q)} \\
& = -mc^2\left(1-\frac{\dot{q}_i\dot{q}^i}{2c^2}\right)\left(1-\frac{V_{qu(k)}}{2mc^2}\right) - V_{(q)} \qquad \text{(A.1.0)} \\
& = -mc^2 + \frac{m\dot{q}_i\dot{q}^i}{2} + \frac{V_{qu(k)}}{2} - V_{(q)} = -mc^2 + L_{cl}
\end{aligned}
$$

And the motion equation read

$$
p_{cl_«} = -\frac{\partial L_{cl}}{\partial \dot{q}^«} = -\frac{\partial\left(-mc^2 + \frac{m\dot{q}_i\dot{q}^i}{2} + \frac{V_{qu(k)}}{2} - V_{(q)}\right)}{\partial \dot{q}^«}, \qquad \text{(A.1.1.a)}
$$

$$
\dot{p}_{cl_«} = -\frac{\partial\left(\frac{m\dot{q}^2}{2} + \frac{V_{qu(k)}}{2} - V_{(q)}\right)}{\partial q^«} = -\frac{\partial\left(\frac{V_{qu(k)}}{2} - V_{(q)}\right)}{\partial q^«} \qquad \text{(A.1.1.b)}
$$

that for the three-dimensional motion equations leads to

$$
p_{cl_i} = \frac{\partial\left(\frac{m\dot{q}_i\dot{q}^i}{2} + \frac{V_{qu(k)}}{2} - V_{(q)}\right)}{\partial \dot{q}_i} = m\dot{q}_i \qquad \text{(A.1.2.a)}
$$

$$
\dot{p}_{cl_i} = \frac{\partial\left(\frac{V_{qu(k)}}{2} - V_{(q)}\right)}{\partial q^«} \qquad \text{(A.1.2.b)}
$$

where equation (A.1.2.b), by using the identity

$$V_{qu} = -\frac{\hbar^2}{m}\frac{\partial_{\text{«}}\partial^{\text{«}} |\cdot|}{|\cdot|} = -\frac{\hbar^2}{m}\frac{\partial_0\partial_0 |\cdot|}{|\cdot|} + \frac{\hbar^2}{m}\frac{\partial_i\partial_i |\cdot|}{|\cdot|},$$

$$= -\frac{\hbar^2}{m}\frac{\partial_0\partial_0 |\cdot|}{|\cdot|} - V_{qu_{cl}}$$

reads

$$\dot{p}_{cl_i} = -\frac{\partial\left(V_{qu_{cl}} + V_{(q)}\right)}{\partial q^{\text{«}}} \tag{A.1.3}$$

That leads to the 3-dimensional hydrodynamic equations that read [11, 29]

$$\dot{q}_i = \frac{\partial H_{cl}}{\partial p_i} = \frac{p_i}{m} = \frac{\partial_i S_{(q,t)}}{m}, \tag{A.1.4.a}$$

$$\dot{p}_i = -\partial_i ( H_{cl} + V_{qu_{cl}} ), \tag{A.1.4.b}$$

coupled to the conservation equation (2.0.3) that in the low velocity limit reads

$$\partial_t |\cdot|^2 + \partial_i (|\cdot|^2 \dot{q}_i) = 0, \tag{A.1.5}$$

where

$$V_{qu_{cl}} = -\frac{\hbar^2}{2m}\frac{\partial_i\partial_i |\cdot|}{|\cdot|}. \tag{A.1.6}$$

For unidimensional systems whose Hamiltonian reads

$$H_{cl} = \frac{p_i p_i}{2m} + V_{(q)}, \tag{A.1.7}$$

by using (A.1.4.a) it follows that the action has the form

$$S = E(t - t_0) + S_{(q)} \tag{A.1.8}$$

so that, by integrating (A.1.4.b), it follows that

$$S_n - S_{n(0)} = \int_{t_0}^{t} dt\left(\frac{p_i p_i}{2m} - V_{(q)} - V_{qu(n)}\right) = \frac{m q_i q_i}{2} - (V_{(q)} + V_{qu(q)n})\int_{t_0}^{t} dt = -E_n(t - t_0). \tag{A.1.9}$$

that with the initial condition

$$\dot{q}_i = \dot{q}_{i(0)} = 0 \tag{A.1.10}$$

reads

$$S_n - S_{n(0)} = -(V_{(q)} + V_{qu(q)n})\int_{t_0}^{t} dt = -E_n(t - t_0) \tag{A.1.11}$$

Moreover, by applying the stationary condition

$$\dot{p}_i = -\partial_i ( H + V_{qu} ) = 0 \tag{A.1.12}$$

for the eigenstates, we obtain the identity

$$V_{(q)} + V_{qu(n)} = constant = E_n \qquad \text{(A.1.13)}$$

that leads to the differential equation for the eigenstates

$$V_{(q)} + V_{qu(n)} = V_{(q)} - (\frac{\hbar^2}{2m})\frac{\partial_i \partial_i | \cdot |}{| \cdot |} = E_n \qquad \text{(A.1.14)}$$

*A.2. Free field*

For a free field (i.e., $V_{(q)} = 0$ ) the differential equation (A.1.14) reads

$$V_{qu} = -(\frac{\hbar^2}{2m}) | \cdot_n |^{-1} \partial_i \partial_i | \cdot_n | = E_n . \qquad \text{(A.2.0)}$$

that admits a solution of type $| \cdot_{n(q)} | = A exp\left(k_i q_i\right)$ with

$$E_n = (\frac{\hbar^2}{2m}) k_i k_i = \frac{p_i p_i}{2m}, \qquad \text{(A.2.1)}$$

leading to a continuum spectrum of energy eigenvalues.

*A.3. The Harmonic oscillator*

For a harmonic oscillator (i.e., $V_{(q)} = \frac{m\mu^2}{2} q^2$ ) the differential equation (A.1.14) reads

$$V_{qu} = -(\frac{\hbar^2}{2m}) | \cdot_n |^{-1} \partial_i \partial_i | \cdot_n | = E_n - \frac{m\mu^2 q^2}{2} . \qquad \text{(A.3.0)}$$

leading to solutions of type

$$| \cdot |_{(q,\, t)} = A_{n(q)}\, exp\left(-aq^2\right), \qquad \text{(A.3.1)}$$

from which we obtain $a = \frac{m\mu}{2\hbar}$ and $A_{n(q)} = H_{n(\frac{m\mu}{2\hbar} q)}$ (where $H_{n(x)}$ represents the *n*-th Hermite polynomial). Therefore, the generic *n*-th eigenstate reads

$$\cdot_{n(q)} = | \cdot |_{(q,\, t)}\, exp\, [\, \frac{i}{\hbar} S_{(q,t)}\, ] = H_{n(\frac{m\mu}{2\hbar} q)}\, exp\left( -\frac{m\mu}{2\hbar} q^2 \right) exp\left( -\frac{iE_n t}{\hbar} \right), \qquad \text{A.3.2)}$$

From (A.3.1-2) it follows that the quantum potential of the n-th eigenstate reads

$$\begin{aligned} V_{qu(n)} &= -(\frac{\hbar^2}{2m}) | \cdot | \partial_i \partial_i | \cdot | \\ &= -\frac{m\mu^2}{2} q^2 + \left[ n \left( \frac{\frac{m\mu}{\hbar} H_{n-1} - 2(n-1) H_{n-2}}{H_n} \right) + \frac{1}{2} \right] \hbar\mu \qquad \text{(A.3.3)} \\ &= -\frac{m\mu^2}{2} q^2 + (n + \frac{1}{2}) \hbar\mu \end{aligned}$$

where it has been used the recurrence formula of the Hermite polynomials $\mathrm{H}_{n+1} = \frac{m\mu}{\hbar} q\mathrm{H}_n - 2n\mathrm{H}_{n-1}$, that by (A.1.14) leads to

$$E_n = V_{qu_n} + V_{(q)} = (n+\frac{1}{2})\hbar\mu \quad \text{(A.3.4)}$$

By making a straightforward check, by using (A.1.1-2, A.3.2-3) the stationary initial conditions

$$\dot{p}_i = -\partial_i(H + V_{qu}) = -\partial_i((n+\frac{1}{2})\hbar\mu) = 0, \quad \text{(A.3.5)}$$

$$\dot{q}_i = \frac{\partial_i S}{m} = 0, \quad \text{(A.3.6)}$$

for the eigenstates.

Generally speaking, the hydrodynamic system of motion equations (A.1.4.a-b) can be applied to many particles problem, or fields, system with the Hamiltonian $H = \sum_{j=1}^{n} \frac{\boldsymbol{p}_{(j)}^2}{2m} + V_{(q_{(1)},\dots q_{(n)})}$ and the quantum potential $V_{qu} = -\frac{\hbar^2}{2m} \sum_{j=1}^{n} \frac{\partial_{q_{i(j)}} \partial_{q_{i(j)}} |\cdot|}{|\cdot|}$.

*A.4. The Free KGE field for particle with null rest mass*

In this case it holds $V_{(q)} = 0, \;\; m = 0$.

The initial and stationary conditions (A.0.6-7) leads to

$$mV_{qu} = -\hbar^2 \frac{\partial_« \partial^« |\cdot|}{|\cdot|} = C = constant \quad \text{(A.4.1)}$$

and

$$-\partial_« S = p_{k_«} = m¥\, \dot{q}_« \sqrt{1 - \frac{V_{qu(k)}}{mc^2}} = C' = constant. \quad \text{(A.4.2)}$$

From (A.4.1) it follows that

$$|\cdot| = A \sin Q_« q^« \quad \text{(A.4.3)}$$

and by posing

$$S = -\hbar k_« q^« = -p_« q^« = -(Et, -p_i q_i), \quad \text{(A.4.4)}$$

from (A.4.2) that

$$\hbar k_« = C' = constant \quad \text{(A.4.5)}$$

leading to

$$mV_{qu} = -\hbar^2 \frac{\partial_« \partial^« |\cdot|}{|\cdot|} = \hbar^2 Q_« Q^« = C'. \quad \text{(A.4.6)}$$

From the above relations the hydrodynamic Hamilton-Jacobi equation (2.0.2) reads

$$\partial_« S\ \partial^« S = -\hbar^2 Q_« Q^« + m^2 c^2 \quad \text{(A.4.7)}$$

that for $m = 0$ leads to

$$k_{«}k^{«} = -\left( k_i k_i - \frac{μ^2}{c^2} \right) = Q_{«}Q^{«} \tag{A.4.8}$$

where, from (A.4.4) it has been used the relation

$$μ = \frac{E}{\hbar}. \tag{A.4.9}$$

Moreover, given that

$$\frac{E}{c^2} = m¥ \sqrt{1 - \frac{\hbar^2 Q_{«}Q^{«}}{m^2c^2}} \tag{A.4.10}$$

in order to have a finite energy for $m = 0$ we must have

$$lim_{m \to 0}\, m¥ = finite \tag{A.4.11}$$

and, hence,

$$|\dot{q}| = c \tag{A.4.12}$$

Moreover, being

$$p_{k_{«}} = \hbar k_{«} = m¥\, \dot{q}_{«} \sqrt{1 - \frac{\hbar^2 Q_{«}Q^{«}}{m^2c^2}} = \frac{E}{c^2}\dot{q}_{«} = C' = constant \tag{A.4.13}$$

$$k_i = \frac{m¥}{\hbar}\, \dot{q}_i \sqrt{1 - \frac{\hbar^2 Q_{«}Q^{«}}{m^2c^2}} = \frac{E}{\hbar c^2}\dot{q}_i = \frac{C'}{\hbar} \tag{A.4.14}$$

by using (A.4.9) and (A.4.12), it follows that

$$\begin{aligned} -k_{«}k^{«} &= \left( k_i k_i - \frac{μ^2}{c^2} \right) = \left( \frac{E}{\hbar c^2} \right)^2 \dot{q}_i \dot{q}_i - \frac{μ^2}{c^2} \\ &= \frac{E^2}{\hbar^2 c^2} - \frac{μ^2}{c^2} = 0 = -Q_{«}Q^{«} \end{aligned} \tag{A.4.15}$$

**Appendix B**

dropped

**Appendix C**

*The covariance of classical law of motion from the equivalence of inertial and gravitational mass*

The covariant form of the equation of motion (2.3.12) in the classical limit (i.e., $V_{qu} = 0$) reads

$$\frac{du_{«}}{ds} - \frac{1}{2}\frac{c}{¥}\frac{\partial g_{ª©}}{\partial q^{«}} u^{ª} u^{©} = +\frac{c}{¥}\frac{\partial ln¥}{\partial q^{«}} \tag{C.1}$$

and in low velocity limit (i,e., $¥ = 1$ and $\frac{\partial ln¥}{\partial q^{«}} = 0$) reduces to

$$m\frac{d\dot{q}_{«}}{dt} = \frac{m}{2}\dot{q}^{\sim}\dot{q}\ \ \frac{\partial g_{\sim}}{\partial q^{«}} \qquad \text{(C.2)}$$

where on the right side there is the effect of the geometry of the space-time that generates the gravitational force. If there is a difference between gravitational mass $m_g$ and inertial one $m_i$ we have to distinguish between that one on the left (that accounts for the inertial effects) and that one on the right that accounts for the gravitational interaction so that (C.2) reads

$$\frac{dm_i\dot{q}_{«}}{dt} = \frac{1}{2}m_g\dot{q}^{\sim}\dot{q}\ \ \frac{\partial g_{\sim}}{\partial q^{«}} \qquad \text{(C.3)}$$

that in the Minkoskian limit of gravity (i.e., $g_{«¬} = \begin{bmatrix} 1+\frac{2¨}{c^2} & 0 & 0 & 0 \\ 0 & -1 & 0 & 0 \\ 0 & 0 & -1 & 0 \\ 0 & 0 & 0 & -1 \end{bmatrix}$ where $¨ = -G\frac{M}{r_{12}}$ [30-31] is the Newtonian gravitational potential) leads to

$$\frac{dm_i\dot{q}_{«}}{dt} = m_g\dot{q}^0\dot{q}^0\frac{1}{c^2}\frac{\partial ¨}{\partial q^{«}} = m_g\frac{\partial ¨}{\partial q^{«}} = m_g(\frac{\partial ¨}{\partial q^{«}}, -\frac{\partial ¨}{\partial q_j}) \qquad \text{(C.4)}$$

and to

$$\frac{dm_i\dot{q}_j}{dt} = -\frac{\partial ¨}{\partial q_j} \qquad \text{(C.5.a)}$$

that for the unidimensional motion $q_{j=1} = r_{12}$ leads to

$$\frac{dm_i\dot{q}}{dt} = -m_g G\frac{M}{q^2} \qquad \text{(C.5.b)}$$

that, if $m_i = m_g$, is independent by the mass of the object leading to the classical motion equation

$$\frac{d\dot{q}}{dt} = -G\frac{M}{q^2} \qquad \text{(C.6)}$$

The covariance of the motion equation (C.1-2) (deriving by (2.3.12)) leads to the equivalence of inertial and gravitational mass in the classical limit.

In this way, the (force of) gravity (i.e., C.1- 3 and C.6)) depends just by the geometry of the space (defined by the metric tensor) but not by the mass of the bodies.

The covariance of motion equation (2.3.12), in the classical limit, makes the geometry of the space a background property that the observer does not perceive directly but as the presence of a field force (generated by the variation of vectors in a curved space-time).

**Appendix D**

*The quantum hydrodynamic motion equation in non-Euclidean space-time for stationary eigenstates (unidimensional systems)*

dropped

**Appendix E**

*The Klein-Gordon equation derived by the covariance of the quantum hydrodynamic equation in non-Euclidean space-time*

By using the constitutive hydrodynamic identities (1.0.1-2), the left part of the covariant KGE (2.3.14) reads

$$\begin{aligned}&\frac{1}{\sqrt{-g}}\partial_\mu\sqrt{-g}\left(g^{\mu\nu}\partial_\nu\psi\right)=\frac{1}{\sqrt{-g}}\partial_\mu\sqrt{-g}\,g^{\mu\nu}e^{\frac{i}{\hbar}S}\left(|\psi|\frac{i}{\hbar}\partial_\nu S+\partial_\nu|\psi|\right)\\&=e^{\frac{i}{\hbar}S}\left(\frac{1}{\sqrt{-g}}\partial_\mu\sqrt{-g}\,g^{\mu\nu}\left(|\psi|\frac{i}{\hbar}\partial_\nu S+\partial_\nu|\psi|\right)+g^{\mu\nu}\left(\frac{i}{\hbar}\partial_\nu|\psi|-\frac{1}{\hbar^2}|\psi|\partial_\nu S\right)\partial_\mu S\right)\end{aligned}\tag{E.1}$$

that, by equating the real and imaginary parts, leads to the system of two hydrodynamic equations

$$\begin{aligned}&-\left(\frac{1}{\sqrt{-g}}\frac{\partial_\mu\sqrt{-g}\,g^{\mu\nu}\partial_\nu|\psi|}{|\psi|}-\frac{1}{\hbar^2}\partial_\nu S g^{\mu\nu}\partial_\mu S\right)\\&=\frac{S_{;\mu}S^{;\mu}}{\hbar^2}-\frac{|\psi|_{;\mu}^{;\mu}}{|\psi|}=\frac{m^2c^2}{\hbar^2}\end{aligned}\tag{E.2}$$

that is the equation (2.3.2) and to

$$\begin{aligned}&\begin{pmatrix}|\psi|g^{\mu\nu}\partial_\mu\left(|\psi|\partial_\nu S\right)+|\psi|^2\partial_\nu S\frac{1}{\sqrt{-g}}\partial_\mu\sqrt{-g}\,g^{\mu\nu}\\+|\psi|g^{\mu\nu}\partial_\nu|\psi|\partial_\mu S\end{pmatrix}\\&=\left(g^{\mu\nu}\frac{\partial_\mu\left(|\psi|^2\partial_\nu S\right)}{|\psi|}+|\psi|^2\partial_\nu S\frac{1}{\sqrt{-g}}\partial_\mu\sqrt{-g}\,g^{\mu\nu}\right)\\&=\frac{1}{\sqrt{-g}}\partial_\mu\left(|\psi|^2\sqrt{-g}\,g^{\mu\nu}\partial_\nu S\right)=0\end{aligned}\tag{E.3}$$

that is the current conservation equation (2.3.3)

**Appendix F**

*The minimum action in the Euclidean hydrodynamic formalism*

By using the Lagrangean approach we are able to write the minimum action principle for the quantum hydrodynamic motion equation.

Since in the hydrodynamic representation the motion equation depends also by the quantum potential and hence by $|\cdot|$ and $\partial_{«}|\cdot|$ (being the quantum potential a second order derivative condition), the problem can be generally described by using the set of independent variables : $x_{«} = (q_{«}, \dot{q}_{«}, |\cdot|, \partial_{«}|\cdot|)$ so that the variation of the hydrodynamic action $S_H = \int\iiint L_H dVdt$ reads (in the following we omit the suffix *H*)

$$
\begin{aligned}
¢S &= \int\iiint\left(|\cdot|^2 \frac{\partial L}{\partial x^{«}} + L\frac{\partial |\cdot|^2}{\partial x^{«}}\right) ¢ x^{«} dVdt \\
&= \int\iiint\left(|\cdot|^2 \frac{\partial L}{\partial x^{«}} ¢ x^{«} + L¢ |\cdot|^2\right) dVdt
\end{aligned}
. \qquad \text{(F.1.1)}
$$

Moreover, being the starting point $q_{«\,a}$ and the final one $q_{«\,b}$ fixed, for two infinitely close pathways on which the variation $¢S$ is calculated, it follows that

$$
\int\iiint L¢ |\cdot|^2 dVdt = 0 \qquad \text{(F.1.2)}
$$

and hence, that

$$
\begin{aligned}
¢S &= \int\iiint |\cdot|^2 \frac{\partial L}{\partial x^{«}} ¢ x^{«} dVdt \\
&= \int\iiint |\cdot|^2 \left( \frac{\partial L}{\partial q^{«}} ¢ q^{«} + \frac{\partial L}{\partial \dot{q}^{«}} ¢ \dot{q}^{«} + \frac{\partial L}{\partial |\cdot|} ¢ |\cdot| + \frac{\partial L}{\partial \partial_{«} |\cdot|} ¢ \partial^{«} |\cdot| \right) dVdt \\
&= \frac{1}{c}\int\iiint |\cdot|^2 \left( \left( \frac{\partial L}{\partial q^{«}} - \frac{d}{dt}\frac{\partial L}{\partial \dot{q}^{«}} \right) ¢ q^{«} + \left( \frac{\partial L}{\partial |\cdot|} - \partial^{«} \frac{\partial L}{\partial \partial_{«} |\cdot|} \right) ¢ |\cdot| \right) d\Omega
\end{aligned}
\qquad \text{(F.1.3)}
$$

where the last equality is obtained by using the identities

$$
\iiint \left[ \left( |\cdot|^2 \frac{\partial L}{\partial \dot{q}^{«}} ¢ q^{«} \right)_{t=t_f} - \left( |\cdot|^2 \frac{\partial L}{\partial \dot{q}^{«}} ¢ q^{«} \right)_{t=t_i} \right] dV = 0 \qquad \text{(F.1.4)}
$$

(because the initial point and the final point are the same for the two pathways) and

$$
\iiint |\cdot|^2 \frac{\partial}{\partial q_{«}} \left( \frac{\partial L}{\partial \partial^{«} |\cdot|} ¢ |\cdot| \right) dV = 0 \qquad \text{(F.1.5)}
$$

that follows since the integration is extended at infinity where the flux of $\frac{\partial L}{\partial \partial^{«} |\cdot|} ¢ |\cdot|$ can be assumed null. Thence, the quantum motion equations (2.1.6-7) can be derived by the generalized minimum action principle

$$¢\left(S - \Delta S\right) = 0 \qquad \text{(F.1.6)}$$

where the variation of the action has to be considered free from the quantum contribution

$$¢\left(\Delta S\right) = -\frac{1}{c}\int\iiint |\cdot|^2 \left(\frac{\partial L}{\partial |\cdot|} - \partial^{«}\frac{\partial L}{\partial \partial_{«}|\cdot|}\right) ¢\,|\cdot|\, d\Omega \qquad \text{(F.1.7)}$$

generated by the term

$$\frac{\partial L}{\partial |\cdot|} - \partial^{«}\frac{\partial L}{\partial \partial^{«}|\cdot|} \qquad \text{(F.1.8)}$$

that derives from the quantum properties of the mass distribution $|\cdot|^2$ that originates by the dependence of the Lagrangean function both from $|\cdot|$ and $\partial^{«}|\cdot|$ contained in the quantum potential.

## Appendix G

*Commutation rules in non-Euclidean space-time*

By applying to a generic vector $B_k$ the commutator $\left[p_{«}, A_{¬}\right]$, where $A_{¬}$ is a generic vector and $p_{«}$ is the operator representing the covariant derivative such as

$$p_{«}(\ )_{k_1....k_i....k_N} \equiv i\hbar\left(\frac{\partial(\ )_{k_1....k_i....k_N}}{\partial q^{«}} - \sum_{i=1}^{N}\Gamma^{m_i}_{k_i«}(\ )_{k_1....m_i....k_N}\right). \qquad \text{(G.1)}$$

it follows that

$$\begin{aligned}\left[p_{«}, A_{¬}\right]B_k &= i\hbar\left(\left(\frac{\partial A_{¬}B_k}{\partial q^{«}} - \Gamma^{m}_{¬«}A_m B_k - \Gamma^{m}_{k«}A_{¬}B_m\right) - A_{¬}\left(\frac{\partial B_k}{\partial q^{«}} - \Gamma^{m}_{k«}B_m\right)\right)\\ &= i\hbar\left(\frac{\partial A_{¬}B_k}{\partial q^{«}} - \Gamma^{m}_{¬«}A_m B_k - \Gamma^{m}_{k«}A_{¬}B_m - A_{¬}\frac{\partial B_k}{\partial q^{«}} - A_{¬}\Gamma^{m}_{k«}B_m\right)\\ &= i\hbar B_k\left(\frac{\partial A_{¬}}{\partial q^{«}} - \Gamma^{m}_{¬«}A_m\right) = B_k A_{¬;«}\end{aligned} \qquad \text{(G.2)}$$

that being $B_k$ undefined leads to

$$\left[p_{«}, A_{¬}\right] = A_{¬;«}. \qquad \text{(G.3)}$$

In the above calculation it has been assumed that the Euclidean space is described by the Minkoskian signature

$$g_{0\,¬«} = ¦_{«¬} = \begin{matrix} 1 & 0 & 0 & 0\\ 0 & -1 & 0 & 0\\ 0 & 0 & -1 & 0\\ 0 & 0 & 0 & -1\end{matrix}. \qquad \text{(G.4)}$$

## Appendix H

Given the generic quantum field state [20]

$$|\,n_k >=|\,n_{k_1}....n_{k_n} >$$

as a collection of $k_n$ harmonic oscillator, the mean squared field value reads

$$< n_k \,||\phi|^2|\, n_k > = < n_k \,|\iint \frac{d^3k'}{(2\pi)^3 2\omega_{k'}} \frac{d^3k''}{(2\pi)^3 2\omega_{k''}} \begin{pmatrix} a(k')a^{\dagger}(k'')e^{-i(k'-k'')q} \\ +a^{\dagger}(k')a(k'')e^{i(k'-k'')q} \\ +a(k')a(k'')e^{-i(k'+k'')q} \\ +a^{\dagger}(k')a^{\dagger}(k'')e^{i(k'+k'')q} \end{pmatrix} | n_k >$$

$$= \iint \frac{d^3k'}{(2\pi)^3 2\omega_{k'}} \frac{d^3k''}{(2\pi)^3 2\omega_{k''}} \begin{pmatrix} < n_k \,| a(k')a^{\dagger}(k'') | n_k > e^{-i(k'-k'')q} \\ + < n_k \,| a^{\dagger}(k')a(k'') | n_k > e^{i(k'-k'')q} \\ + < n_k \,| a(k')a(k'') | n_k > e^{-i(k'+k'')q} \\ + < n_k \,| a^{\dagger}(k')a^{\dagger}(k'') | n_k > e^{i(k'+k'')q} \end{pmatrix} \quad \text{(H.1)}$$

That, by using the commutation relations (5.0.12-14), leads to

$$= \iint \frac{d^3k'}{(2\pi)^3 2\omega_{k'}} \frac{d^3k''}{(2\pi)^3 2\omega_{k''}} \sqrt{2\omega_{k'}}\sqrt{2\omega_{k''}} < n_k | n_k > \left( (n_{k'}+1)\delta_{k'-k''} e^{-i(k'-k'')q} + n_{k'}\delta_{k'-k''} e^{i(k'-k'')q} \right)$$

$$= < n_k | n_k > \int \frac{d^3k}{(2\pi)^3 2\omega_k} (2n_k+1) = \frac{1}{c^2} \int \frac{kdk}{(2\pi)^2} (2n_k+1) \quad , \quad \text{(H.2)}$$

$$= \frac{1}{(2\pi)^2 c^2} \int^{\omega_{k_{max}}} (2n_k+1) \sqrt{\frac{\omega_k^{\,2}}{c^2} - \frac{m^2c^4}{c^2\hbar^2}\left(1-\frac{V_{qu(k)}}{mc^2}\right)}\, d\omega_k$$

that for the fundamental state, leads to

$$< 0_k \,||\phi|^2|\, 0_k > = \frac{1}{(2\pi)^2 c^2} \int^{\omega_{k_{max}}} \sqrt{\frac{\omega_k^{\,2}}{c^2} - \frac{m^2c^4}{c^2\hbar^2}\left(1-\frac{V_{qu(k)}}{mc^2}\right)}\, d\omega_k \quad \text{(H.3)}$$

and by (6.0.12), to

$$< 0_k \,||\phi|^2|\, 0_k > = \frac{2\pi}{(2\pi)^3 c^2} \int^{\omega_{k_{max}}} \sqrt{\frac{\omega_k^{\,2}}{c^2} - \frac{m^2c^4}{c^2\hbar^2}}\, d\omega_k \,. \quad \text{(H.4)}$$

As far as it concerns the zero-point vacuum energy, we obtain

$$< 0_k | H_0 | 0_k > = \frac{\hbar}{2} \int \frac{d^3k}{(2\pi)^3} < 0_k \,|\left( a_0^{\dagger}(k)a_0(k) + a_0(k)a_0^{\dagger}(k) \right)|\, 0_k >$$

$$= \hbar \int \frac{d^3k}{(2\pi)^3} < 0_k \,| a_0^{\dagger}(k)a_0(k) + \frac{\omega_k}{2} |\, 0_k >$$

$$= \frac{\hbar}{2} < 0_k || 0_k > \int \frac{d^3k}{(2\pi)^3} \omega_k \quad \text{(H.5)}$$

$$= \frac{\hbar}{2(2\pi)^2 c^2} \int^{\omega_{k_{max}}} \omega_k^{\,2} \sqrt{\frac{\omega_k^{\,2}}{c^2} - \frac{m^2c^4}{c^2\hbar^2}\left(1-\frac{V_{qu(k)}}{mc^2}\right)}\, d\omega_k$$

that by (6.0.12) leads to

$$< 0_k \,|\, H_0 \,|\, 0_k > = \frac{\hbar}{2(2\text{®})^2 c^2} \int^{\mu_{k_{max}}} \mu_k{}^2 \sqrt{\frac{\mu_k{}^2}{c^2} - \frac{m^2 c^4}{c^2 \hbar^2}} d\mu_k \qquad \text{(H.6)}$$

**Appendix I**

dropped

**Appendix J**

*Evaluation of the mean CPTD on the cosmological scale*

If in the classical Einstein equation the "cosmological" constant $\Lambda$ (a proportionality factor multiplying $g_{\text{«¬}}$ ) cannot be function of the space (if we do not want to alter the classical equations of motion) on the contrary, in the GE (3.2.18) the cosmological energy-impulse tensor density $\Lambda g_{\text{«¬}}$ is both a function of the mass distribution density $|\cdot|^2$ and of the quantum potential.

As shown by (6.0.15), the CC is different from zero only in the regions of space where the mass is very concentrated (e.g., $\frac{\partial|\cdot|}{\partial q^{\text{«}}} \neq 0$ ) while it is null in the vacuum given that if $\frac{\partial|\cdot|}{\partial q^{\text{«}}} = 0$ also $V_{qu} = 0$.

Thence, the effect on cosmological scale on the mass of galaxies of the term $\Lambda g_{\text{«¬}}$ given by the mean value $< \Lambda >$ over all the volume of the galactic space $V_g$, that reads

$$< \frac{\Lambda}{c^2} > = \frac{1}{V_g} \iiint_{V_g} \frac{\Lambda}{c^2} dV \,, \qquad \text{(J.1)}$$

takes contributions just from the volume near the particles of matter.

In order to fully evaluate expression (J.1), we should introduce the electromagnetic, the strong and the weak forces, and the Higgs boson interaction that give rise to the family of elemental particles (e.g., of the standard model [32]).

For the model developed so far, we only have a scalar field coupled to the gravitational force so that the localization of the mass can only come from gravity whose stable localized states are those of the black holes (BH) with a mass larger than the Planck mass [15]. In order to make an evaluation of (J.1) by using this oversimplified model, we have to represent all the mass of the universe (baryonic matter, dark matter and dark energy respectively) by particles of BHs of Plancknian mass (PBH) with the volume of a sphere of Planck radius $\text{¢}V_{pBH} \approx 2\ 10^{-105} m^3$. Thence, by posing that

$$< \frac{\Lambda}{c^2} > \approx \frac{1}{V_g} \begin{pmatrix} < \frac{\Lambda}{c^2} >_{barionic} + < \frac{\Lambda}{c^2} >_{dark\ m} \\ + < \frac{\Lambda}{c^2} >_{dark\ En} \end{pmatrix} \text{¢}V_{pBH} \qquad \text{(J.2)}$$

where $V_g$ is the galactic volume considered, and, by utilizing (3.2.22), it follows that

$$< \frac{\Lambda}{c^2} >_{barionic} + < \frac{\Lambda}{c^2} >_{dark\ m} \approx \frac{m\text{¢}V_{pBH}}{V_g} < \frac{|\cdot|^2_{pBH(m)}}{\text{¥}} \left(1 - \sqrt{1 - \frac{V_{qu}}{mc^2}}\right) > \qquad \text{(J.3)}$$

where $m = m_{barionic} + m_{dark\ m}$ , where $m_{barionic}\text{¢}V_{pBH}$ and $m_{dark\ m}\text{¢}V_{pBH}$ are the baryonic and dark matter mass, respectively, of each PBH).

The zero-point energy density of the vacuum (that is the consequence of field quantization) is taken into account by the term of dark energy.

In figure 1, it is depicted the discretization used to evaluate (J.3). Similarly to the quantum loop gravity, the vacuum is assumed composed of cells of volume $\text{¢}V_{pBH} \approx 2\ 10^{-105} m^3$ whose radius is of order of the Planck length magnitude. Some of them contain the localized mass $m$ of the PBH of baryonic and dark matter.

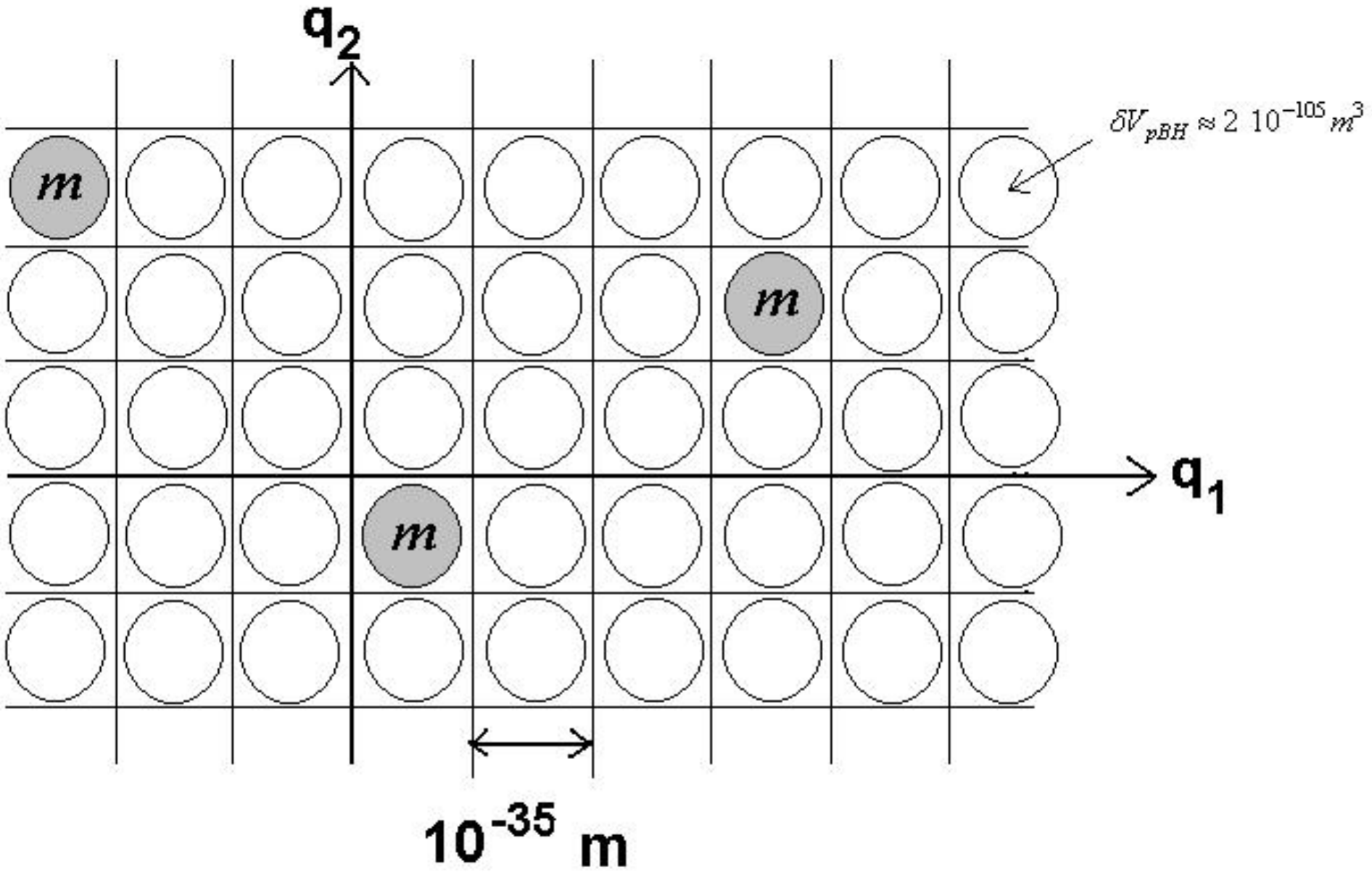

FIG. 1

**Figure 1.** The (baryonic and dark) mass distribution in the Plancknian cells of vacuum

Moreover, we introduce the zero-point energy density of the vacuum $E_0$ by adding it, or the equivalent mass, $m_{dark\ En} = \frac{E_0}{c^2}$, to each cell of the space, to obtain

$$< \frac{\Lambda}{c^2} >_{dark\ En} \approx \frac{\frac{E_0}{c^2}\text{¢}V_{pBH}}{V_g} < \frac{|\cdot|^2_{dark\ En}}{\text{¥}} \left( 1 - \sqrt{1 - \frac{V_{qu}}{mc^2}} \right) > \quad . \qquad (J.4)$$

After that, we obtain the set up represented in figure 2

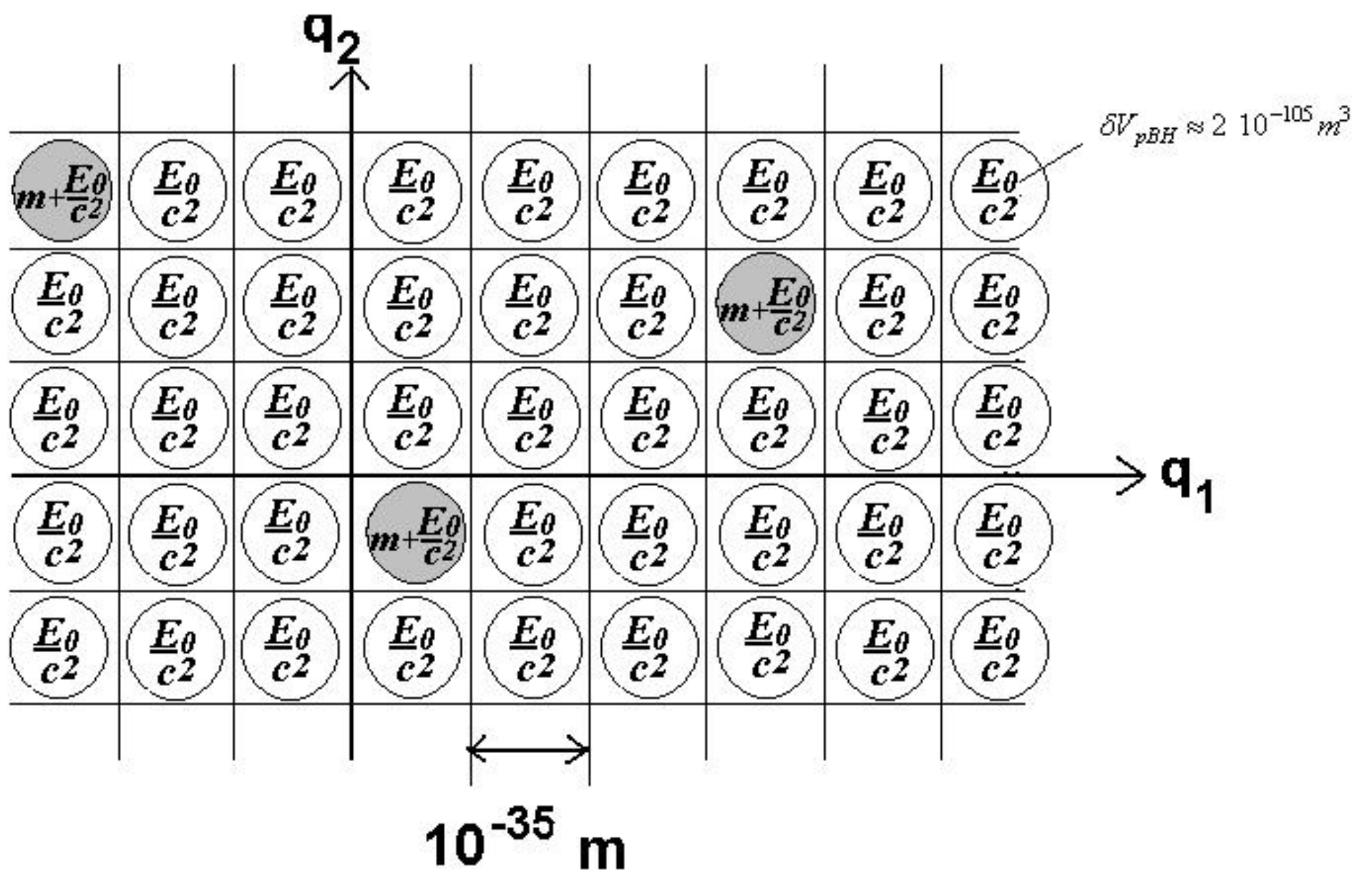

FIG. 2

**Figure 2.** The mass (baryonic and dark ones) and dark energy distribution in the Plancknian cells of vacuum

Moreover, given that the constant mass distribution $|\cdot|_{tot}$ in the vacuum (far from the PBHs mass) does not give contribution both to the quantum potential (6.0.12) and, hence, to the cosmological constant (6.0.9) it follows that in (J.3) we can make the approximation that only the places where there are both the baryonic and the dark matter give a not null zero point energy contribution .

Thence, we can assume that the number of PBH per volume $|\cdot|^2_{dark\ En}$, that account for the dark energy, are equal to the number of PBH per volume accounting for the matter $|\cdot|^2_m$.

Thence, from (J.3) It follows that

$$< \frac{\Lambda}{c^2} > \approx \frac{m_{tot} ¢ V_{pBH}}{V_g} < \frac{|\cdot|^2_{pBH}}{¥} \left(1-\sqrt{1-\frac{V_{qu}}{mc^2}}\right) > \qquad (J.5)$$

where $m_{tot} = m + \frac{E_0}{c^2}$ .

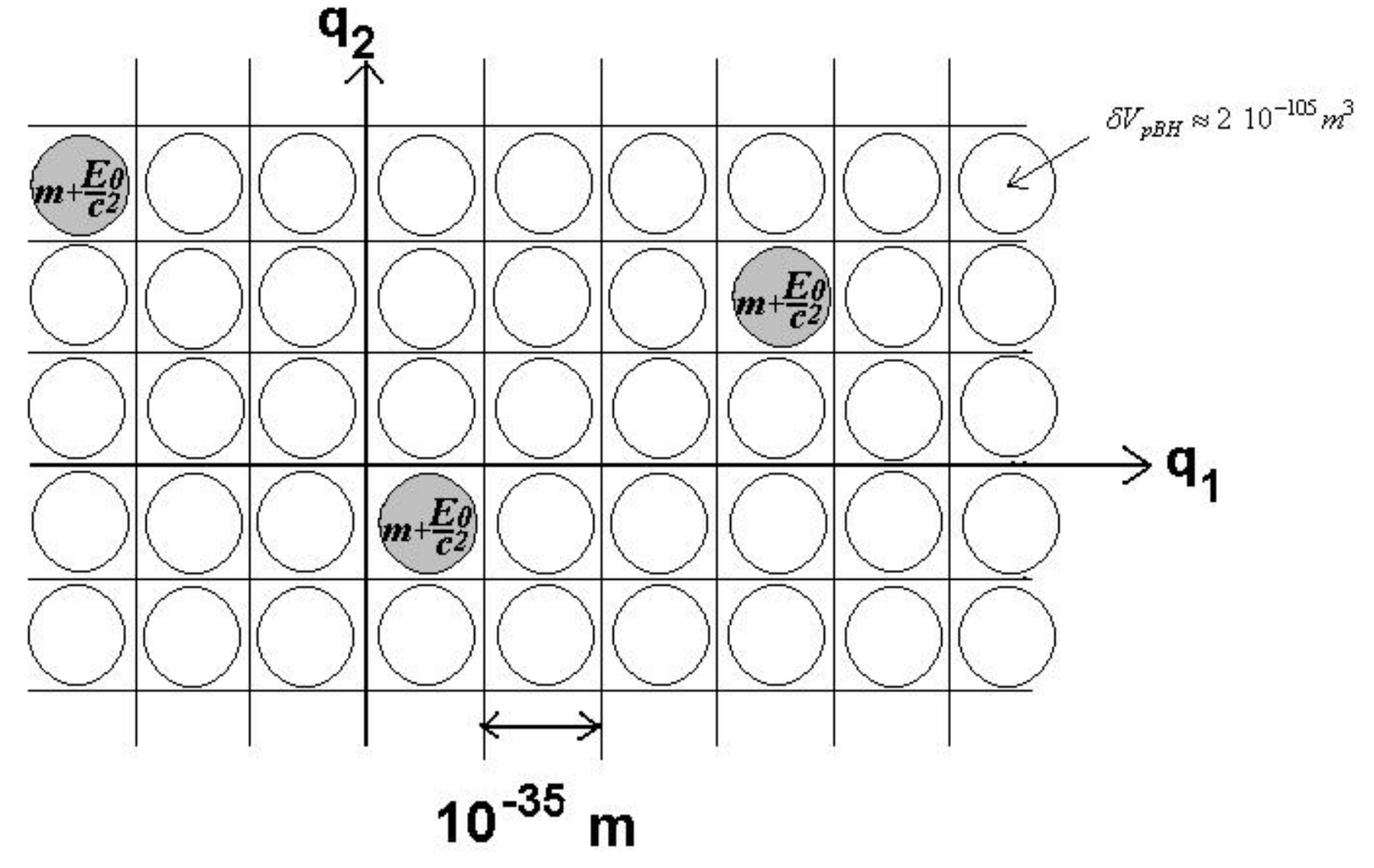

FIG. 3

**Figure 3.** The approximated mass (baryonic and dark ones) and dark energy distribution, in the Plancknian cells of vacuum, that give contributions to the cosmological energy-impulse tensor

Moreover, by posing both that

$$< \frac{|\cdot|^2_{pBH}}{¥}\left(1-\sqrt{1-\frac{V_{qu}}{mc^2}}\right) > \approx < |\cdot|^2_{pBH} > < \frac{1}{¥}\left(1-\sqrt{1-\frac{V_{qu}}{mc^2}}\right) > \qquad \text{(J.6)}$$

and that the mean volume, in the galactic space, available per PBH mass [1] (that for the baryonic mass, 4% of the total mass, is of about eight hydrogen atoms per cubic meters) reads

$$V_{mean} = \frac{V_g}{< |\cdot|^2_{pBH} >} \approx 2\ 10^5 m^3, \qquad \text{(J.7)}$$

it follows that

$$< \frac{\Lambda}{c^2} > \approx \frac{\left(m_{b+d} + \frac{E_0}{c^2}\right) ¢ V_{pBH}}{V_{mean}} < \frac{1}{¥}\left(1-\sqrt{1-\frac{V_{qu}}{mc^2}}\right) >, \qquad \text{(J.8)}$$

Furthermore, in order to evaluate the order of magnitude of $< \frac{\Lambda}{c^2} >$, we introduces the experimental data both for the baryonic and dark matter density $\Omega_m$ and for the dark energy density $\Omega_{dark\ En}$ in the expression (J.8) to obtain

$$< \frac{\Lambda}{c^2} > \approx \frac{\left( m_{b+d} + \frac{E_0}{c^2} \right) ¢ V_{pBH}}{V_{mean}} \frac{1}{< ¥ >} < \left( 1 - \sqrt{1 - \frac{V_{qu}}{mc^2}} \right) >$$

$$\approx \left( \Omega_m + \Omega_{dark\ En} \right) \frac{1}{< ¥ >} \left( 1 - \sqrt{1 - \frac{< V_{qu} >}{mc^2}} \right) \qquad (J.9)$$

where [1]

$$\Omega_m = \frac{m_{b+d} ¢ V_{pBH}}{V_{mean}} \cong 2{,}6\ 10^{-27} kg / m^3, \qquad (J.10)$$

and where

$$\Omega_{dark\ En} = \frac{\frac{E_0}{c^2} ¢ V_{pBH}}{V_{mean}} ¢ V_{pBH} \cong 7{,}4\ 10^{-27} kg / m^3 . \qquad (J.11)$$

As far as it concerns

$$< ¥ > \approx \frac{1}{\sqrt{1 - \frac{< \dot{q}^2 >}{c^2}}} \qquad (J.12)$$

we can approximate it of about the unity for low velocity massive particles (e.g., moving at speed lower than $\frac{1}{3} c$ ) as shown below

$$< ¥ > \quad \approx < \frac{1}{\sqrt{1 - \frac{\left( \frac{c}{6} \right)^2}{c^2}}} \cong \frac{1}{\sqrt{1 - \frac{1}{36}}} \cong 1 + \frac{1}{72} = 1{,}013\bar{8} . \qquad (J.13)$$

As far as it concerns $< V_{qu} >$, we can evaluate it for a Schwarzschild radial symmetric BH.

Given the metric tensors in the spherical coordinates

$$g_{00} = e^{¬} ; g_{11} = -e^{ª} ; g_{22} = -r^2 ;$$

$$g_{33} = -r^2 sin^2 — ; \ \sqrt{-g} = | e^{\frac{ª + ¬}{2}} r^4 sin^2 — |^{-1} , \qquad (J.14)$$

for the Schwarzschild BH, they read [15, 30]

$$g_{11} = -e^{ª} = -g_{00}{}^{-1} \qquad (J.15)$$

$$g_{00} = e^{¬} = e^{-ª} = -\left( 1 - \frac{R_g}{r} \right) \qquad (J.16)$$

$$g = -r^4 sin^2 \mathrm{g} . \qquad (J.17)$$

By introducing (J.15-17) in the hydrodynamic equation of motion for eigenstates (2.3.13) with the stationary conditions

$$\frac{du_«}{dt} = 0 \qquad (J.18)$$

$u_{«} = (1,0,0,0)$ ,

(J.19)

it follows that

$$0 = \frac{1}{2}\frac{c}{¥} u^{a} u^{©} \frac{\partial g_{a©}}{\partial q^{«}} + \frac{c}{¥}\frac{\partial ln\sqrt{1-\frac{V_{qu(n)}}{mc^{2}}}}{\partial q^{«}} + \frac{c}{¥}\frac{\partial ln ¥}{\partial q^{«}} - u_{«}\frac{d}{dt} ln\sqrt{1-\frac{V_{qu}}{mc^{2}}} \qquad (J.20)$$

and, finally, (see Appendix H) that

$$\left(1-\frac{V_{qu}}{mc^{2}}\right) = g_{00}{}^{-3}, \qquad (J.21)$$

that

$$1-\left(1-\frac{R_{g}}{r}\right)^{-3} = \frac{V_{qu}}{mc^{2}} \qquad (J.22)$$

and that

$$< V_{qu} >= mc^{2} < \left(1-\left(1-\frac{R_{g}}{r}\right)^{-3}\right) >\cong mc^{2} < \left(1-\left(\frac{r}{R_{g}}\right)^{3}\right) > . \qquad (J.23)$$

Moreover, since the mass of a black hole is almost concentrated in $r << R_{g}$ [15, 30], it follows that

$$< V_{qu} >_{(n)} \approx mc^{2} \qquad \forall n > 0, \qquad (J.24)$$

that

$$< \frac{\Lambda}{c^{2}} >\cong 0{,}9863\left(\Omega_{m} + \Omega_{dark\ En}\right) \qquad (J.25)$$

and, hence, by using (J.11) that

$$E_{0} \approx \Omega_{dark\ En}\ c^{2}\frac{V_{mean}}{¢V_{bBH}} \approx 8\ 10^{98} J/m^{3} \qquad (J.26).$$

Moreover, if we consider the zero-point energy density $E_{0}$ of the vacuum [33] in QCD

$$E_{0} \cong \frac{\hbar}{8®^{2}c^{3}}\mu^{4}{}_{max} \qquad (J.27)$$

from (J.26) it follows that

$$\hbar\mu_{max} \approx 3\ 10^{18} GeV \qquad (J.28)$$

that is in good agreement with the maximum allowed Plancknian cut-off frequency

$$\hbar\mu_{max} = \frac{\hbar c}{l_{p}} = m_{p}c^{2} = c^{2}\sqrt{\frac{\hbar c}{16®G}} \approx 5\ 10^{18} GeV . \qquad (J.29)$$

where $l_{p}$ is the Planck length.